\documentclass[a4paper,fleqn,usenatbib,useAMS]{mnras}
 
\usepackage{mathptmx}
\usepackage{txfonts}

\usepackage[T1]{fontenc}
\usepackage{ae,aecompl}

\usepackage{graphicx}	
\usepackage{epsfig}

\usepackage{natbib}
\usepackage{footnote}
\usepackage{lscape}
\usepackage{longtable}
\usepackage{supertabular}
\usepackage{caption}
\usepackage{subfig}
\usepackage{txfonts}

\title[Sublimation of icy grains in 67P/C-G with OSIRIS.]{Sublimation of icy aggregates in the coma of comet 67P/Churyumov-Gerasimenko detected with the OSIRIS cameras onboard Rosetta.}

\author[A. Gicquel et al.]{
\parbox{\textwidth}{
\begin{normalsize}
A. Gicquel$^{1}$\thanks{E-mail: gicquel@mps.mpg.de},
J.-B. Vincent$^{1}$, J. Agarwal$^{1}$, M. F. A'Hearn$^{2}$, I. Bertini$^{3}$, D. Bodewits$^{2}$, H. Sierks$^{1}$, Z.-Y. Lin$^{4}$, C. Barbieri$^{5}$, P. L. Lamy$^{6}$, R. Rodrigo$^{7,8}$, D. Koschny$^{9}$, H. Rickman$^{10,11}$, H. U. Keller$^{12}$, M. A. Barucci$^{13}$, J.-L. Bertaux$^{14}$, S. Besse$^{9}$, G. Cremonese$^{15}$, V. Da Deppo$^{16}$, B. Davidsson$^{10}$, S. Debei$^{17}$, J. Deller$^{1}$, M. De Cecco$^{18}$, E. Frattin$^{15}$, M. R. El-Maarry$^{19}$, S. Fornasier$^{13}$, M. Fulle$^{20}$, O. Groussin$^{21}$, P. J. Guti\'errez$^{22}$, P. Guti\'errez-Marquez$^{1}$, C. G\"uttler$^{1}$, S. H\"ofner$^{1,12}$, M. Hofmann$^{1}$, X. Hu$^{1}$, S. F. Hviid$^{23}$, W.-H. Ip$^{4}$, L. Jorda$^{21}$, J. Knollenberg$^{23}$, G. Kovacs$^{1,24}$, J.-R. Kramm$^{1}$, E. K\"uhrt$^{23}$, M. K\"uppers$^{25}$, L. M. Lara$^{22}$, M. Lazzarin$^{5}$, J. J. Lopez Moreno$^{22}$, S. Lowry$^{26}$, F. Marzari$^{5}$, N. Masoumzadeh$^{1}$, M. Massironi$^{3}$, F. Moreno$^{22}$, S. Mottola$^{23}$, G. Naletto$^{27,2,16}$, N. Oklay$^{1}$, M. Pajola$^{3}$, A. Pommerol$^{19}$, F. Preusker$^{23}$, F. Scholten$^{23}$, X. Shi$^{1}$, N. Thomas$^{19}$, I. Toth$^{27,6}$, C. Tubiana,$^{1}$
\end{normalsize}} 
\\~\\
\parbox{\textwidth}{
$^{1}$Max-Planck Instit\"ut fur Sonnensystemforschung, Justus-von-Liebig-Weg 3, 37077 G\"ottingen, Germany; $^{2}$Department for Astronomy, University of Maryland, College Park, MD 20742-2421, USA; $^{3}$Centro di Ateneo di Studi ed Attivit\'{a} Spaziali "Giuseppe Colombo" (CISAS), University of Padova, Via Venezia 15, 35131 Padova, Italy; $^{4}$Institute for Space Science, National Central University, 32054 Chung-Li, Taiwan; $^{5}$Department of Physics and Astronomy "G. Galilei", University of Padova, Vic. Osservatorio 3, 35122 Padova, Italy; $^{6}$Laboratoire d'Astrophysique de Marseille, UMR 7326, 13388, $\&$ Aix-Marseille Universit\'{e}, Marseille, France; $^{7}$Centro de Astrobiologia (INTA-CSIC), European Space Agency, European Space Astronomy Centre (ESAC), P.O. Box 78, E-28691 Villanueva de la Canada, Madrid, Spain; $^{8}$International Space Science Institute, Hallerstrasse 6, 3012 Bern, Switzerland; $^{9}$Research and Scientific Support Department, European Space Agency, 2201 Noordwijk, The Netherlands; $^{10}$Department of Physics and Astronomy, Uppsala University, Box 516, 75120 Uppsala, Sweden; $^{11}$PAS Space Research Center, Bartycka 18A, 00716 Warszawa, Poland;  $^{12}$Institute for Geophysics and Extraterrestrial Physics, TU Braunschweig, 38106 Braunschweig, Germany; $^{13}$LESIA, Observatoire de Paris, CNRS, UPMC Univ Paris 06, Univ. Paris-Diderot, 5 Place J. Janssen, 92195 Meudon Pricipal Cedex, France; $^{14}$LATMOS, CNRS/UVSQ/IPSL, 11 Boulevard d'Alembert, 78280 Guyancourt, France; $^{15}$INAF Osservatorio Astronomico di Padova, Vicolo dell'Osservatorio 5, 35122 Padova, Italy; $^{16}$CNR-IFN UOS Padova LUXOR, Via Trasea 7, 35131 Padova, Italy; $^{17}$Department of Industrial Engineering University of Padova Via Venezia, 1, 35131 Padova, Italy; $^{18}$University of Trento, via Sommarive, 9, Trento, Italy; $^{19}$Physikalisches Institut, Sidlerstrasse 5, University of Bern, CH-3012 Bern, Switzerland; $^{20}$INAF - Osservatorio Astronomico di Trieste, via Tiepolo 11, 34143 Trieste, Italy; $^{21}$Aix Marseille Universit\'e, CNRS, LAM (Laboratoire d'Astro-physique de Marseille) UMR 7326, 13388, Marseille, France; $^{22}$Instituto de Astrofisica de Andalucia-CSIC, Glorieta de la Astronomia, 18008 Granada, Spain; $^{23}$Institute of Planetary Research, DLR, Rutherfordstrasse 2, 12489 Berlin, Germany; $^{24}$Budapest University of Technology and Economics, Department of Mechatronics, Optics and Engineering Informatics, Muegyetem rkp 3, Budapest, Hungary; $^{25}$ESA/ESAC, PO Box 78, 28691 Villanueva de la Ca\~nada, Spain; $^{26}$Centre for Astrophysics and Planetary Science, School of Physical Sciences, The University of Kent, Canterbury CT2 7NH, United Kingdom; $^{27}$Department of Information Engineering, University of Padova, Via Gradenigo 6/B, 35131 Padova, Italy; $^{27}$Observatory of the Hungarian Academy of Sciences, PO Box 67, 1525 Budapest, Hungary.
}}

\date{Accepted 2016 August 19. Received 2016 August 19; in original form 2016 May 04}

\pubyear{2016}

\begin{document}
\label{firstpage}
\pagerange{\pageref{firstpage}--\pageref{lastpage}}
\maketitle

\begin{abstract}
Beginning in March 2014, the OSIRIS (Optical, Spectroscopic, and Infrared Remote Imaging System) cameras began capturing images of the nucleus and coma (gas and dust) of comet 67P/Churyumov-Gerasimenko using both the wide angle camera (WAC) and the narrow angle camera (NAC). The many observations taken since July of 2014 have been used to study the morphology, location, and temporal variation of the comet's dust jets. We analyzed the dust monitoring observations shortly after the southern vernal equinox  on May 30 and 31, 2015 with the WAC at the heliocentric distance $\rm R_h$ = 1.53 AU, where it is possible to observe that the jet rotates with the nucleus. We found that the decline of brightness as a function of the distance of the jet is much steeper than the background coma, which is a first indication of sublimation. We adapted a model of sublimation of icy aggregates and studied the effect as a function of the physical properties of the aggregates (composition and size). The major finding of this article was that through the sublimation of the aggregates of dirty grains (radius $a$ between 5$\rm \mu$m and 50$\rm \mu$m) we were able to completely reproduce the radial brightness profile of a jet beyond 4 km from the nucleus. To reproduce the data we needed to inject a number of aggregates between 8.5 $\times$ $10^{13}$ and 8.5 $\times$ $10^{10}$ for $a$ = 5$\mu$m and 50$\mu$m respectively, or an initial mass of $\rm H_2O$ ice around 22kg.
\end{abstract}



\section{Introduction}

The composition of cometary nuclei (mainly ices, silicate dust and organics) is known primarily from observations of their comae \citep{AHearn_2011, Hanner_Zolensky_2010, Lisse_2006}. The structure of a cometary coma is often expressed in terms of a uniform, spherically-symmetric outflow. The volatile species are either parent species, sublimated directly from the nucleus, or daughter species, produced in the coma \citep{Haser_1957}. The picture is complicated by chemical processes, temporal evolution and the sublimation of icy aggregates. In situ spacecraft observations of several comets have allowed their nuclei to be imaged and the gas and dust properties of their coma to be studied in great detail. More specifically, the sublimation of icy grains has been detected in the ejecta of comet 9P/Tempel 1 after the impact with the Deep Impact spacecraft \citep{Gicquel_2012} and in the coma of comet 103P/Hartley 2 during the Deep Impact eXtended Investigation (DIXI) \citep{Protopapa_2015}.\\

Since its discovery in 1969, comet 67P/Churyumov-Gerasimenko (67P) has been observed from ground-based observatories during its 7 apparitions. From 2008-2009, i.e. during  its last perihelion, \cite{Lara_2015}, \cite{Tozzi_2011}, and \cite{Vincent_2013} monitored a highly anisotropic outgassing where many jets were detected. The ESA (European Space Agency) Rosetta spacecraft was launched on March 2, 2004, to reach comet 67P/Churyumov-Gerasimenko (67P). Since March 2014, images of the nucleus and the coma (gas and dust) of the comet have been acquired by the OSIRIS (Optical, Spectroscopic, and Infrared Remote Imaging System) camera system \citep{Keller_2007} using both the wide angle camera (WAC) and the narrow angle camera (NAC). The orbiter is expected to remain in close proximity to the nucleus of the comet until post-perihelion in September 2016. To date, more than 2 years of continuous, close-up observations of the gas and dust coma have been obtained and have helped to characterize the evolution of comet gas and dust activity (outgassing effect) during the approach to the Sun. For the first time, we were able to follow the development of a comet's coma from a close distance, and we were able study the dust-gas interaction searching for indications of ongoing sublimation of icy aggregates. Many studies have shown evidence of sublimation on 67P \citep{Lin_2015, Hofstadter_2016}\\

\begin{figure}
\centering
\includegraphics[scale=0.13]{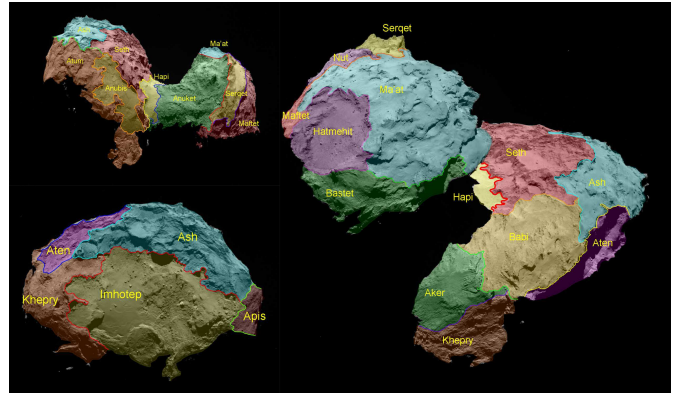}
  \vspace{0.3in}
\caption{Regional definitions based on large-scale unit boundaries. Figure reproduced from Thomas et al. (2015) with permission.}
\label{Map_Thomas}
\end{figure}

From July 2014 onwards, the morphology, the location and the temporal variation of the comet's jets have been observed with the OSIRIS cameras. \cite{Lara_2015} analyzed three OSIRIS data sets in July and August 2014 and found active regions in Hapi, Hathor, Anuket, and Aten (highly spatially resolved). The nucleus map (Shown in Figure \ref{Map_Thomas}) which indicated the regions, was taken from \cite{Thomas_2015}. \cite{Lin_2015} further studied OSIRIS sequences between August and September 2014. During the two months of observations, the cometary distance ranged from 98.7 to 106.1 km and the spatial resolution of the nucleus and the jets was high (< 10 m $\rm pixel^{-1}$). These jets (mostly originating from the Hapi region) did not appear to be associated with specific morphologies but followed insolation \citep{Keller_2015}. \cite{Lin_2016} (May to August 2015) and \cite{Vincent_2016} (August 2014 to May 2015) showed that the jets appeared to rotate with the nucleus and followed the solar illumination. Over the above period (July 2014 to August 2015), the comet passed from northern summer to southern summer with the equinox on May 10 2015. \cite{Vincent_2016} noted that some jets observed in the northern summer on 67P appeared to be morphology-dependent. \\

The present work analyzes if the sublimation of icy aggregates would explain the brightness profile of the jets observed with the OSIRIS data. We analyze the observations acquired at the end of May 2015 with the WAC very shortly after the southern vernal equinox. We study the brightness distribution of the jets ($B$ [W $\rm m^{-2}$ $\rm nm^{-1}$ $\rm sr^{-1}$]) as a function of the distance from the nucleus ($D$ [km]). First, we present the observations obtained with the OSIRIS cameras. We describe the method we used to obtain the photometric profile of the dust jet from the WAC images. The temporal variation of the jet after one rotation of the nucleus is studied and the location of the jet is determined. Additionally, we describe the model of sublimation of icy aggregates which is adapted from the model of sublimation described in \cite{Gicquel_2012}. We follow the evolution of aggregates once they are ejected into the coma until they sublimate completely. Given an aggregate velocity, we determine at which radial distance from the nucleus the sublimation occurs. Finally, we compare the measured radial profile of the jet with the synthetic radial profile of the sublimating icy aggregates.

\section{Observations and data analysis}

\subsection{Data with the OSIRIS Cameras}
\label{Obs_Section}
The OSIRIS cameras, composed of the WAC and NAC, were dedicated to mapping the nucleus of comet 67P and to characterizing the evolution of the comet's gas and dust \citep{Keller_2007}. The WAC (230 - 750 nm) was principally used to study the coma of dust and gas, while the NAC (250 - 1000 nm) was mainly used to investigate the structure of the nucleus. We typically monitored activity of gas and dust with the WAC approximately once every two weeks at heliocentric distances greater than 2 AU and once per week afterward. The nominal sequence for these observations had a set of observations once per hour for a full comet rotation \citep[][12.4h]{Tubiana_2015}. The rotation period of the nucleus decreased by 0.36h during its last perihelion \citep{Mottola_2014, Sierks_2015}. As both Rosetta and the comet approached Earth, the data volume available increased, and we updated the observational sequence to one set of observations every 20 min, for 14h. This provided coverage of the diurnal and seasonal evolution of the coma. \\

Rosetta started to observe 67P, during the long northern summer, while the southern hemisphere was almost in total darkness. In March 2015, as the equinox (10 May 2015) was approaching, we started to observe new regions becoming active close to the equator in the Nut, Serqet, Ma'at and Imhotep regions \citep{Vincent_2016, Lin_2016}. These regions, shown in Figure \ref{Map_Thomas}, had been in darkness, so we expected the release of fresh material during the equinox, because sublimation is strongly correlated with the local insolation \citep{Keller_2015, Schloerb_2015}. The reader is referred to Section \ref{Jet_Location} for a further explanation of the source location of the jet (Imhotep region), but we point out here that we chose to analyze a jet for which the latitude of the base of the jet was almost identical to the sub-solar latitude, i.e., the sun passes directly overhead at noon. Thus the base of the jet receives the maximum duration of sunlight in simple geometries and in this case the geometry allowed the jet-base to be illuminated for a long time each day. Therefore, we chose the dust monitoring observations on May 30 and 31, 2015 utilizing the WAC (MTP016/STP058 DUST MON 005) to compare with our model. \\
 
\begin{figure*}
  \centering
  \subfloat[2015-05-30T15:43:38]{\label{15.43}\includegraphics[scale=0.252]{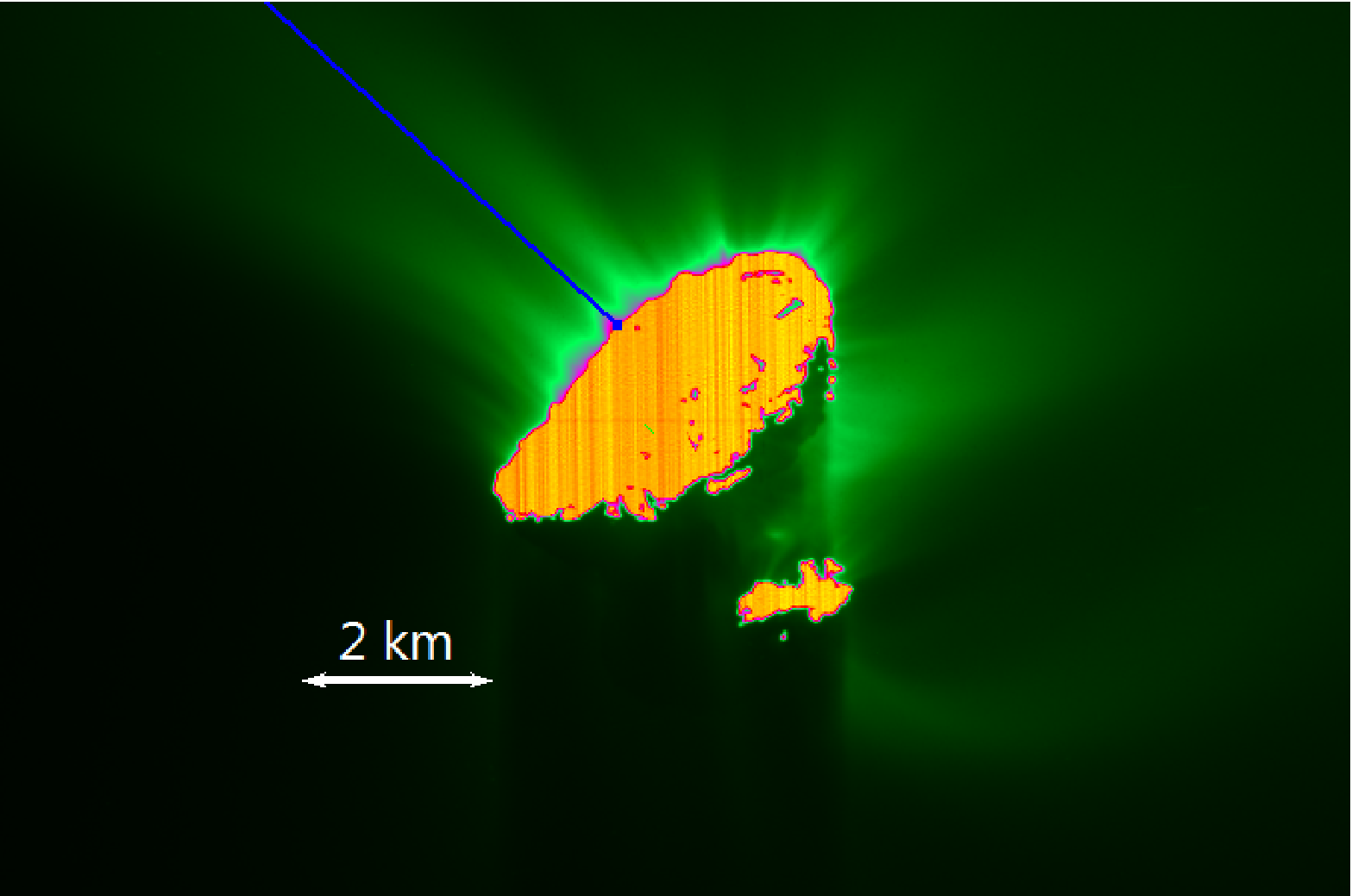}} 
  \subfloat[2015-05-30T16:13:39]{\label{16.13}\includegraphics[scale=0.25]{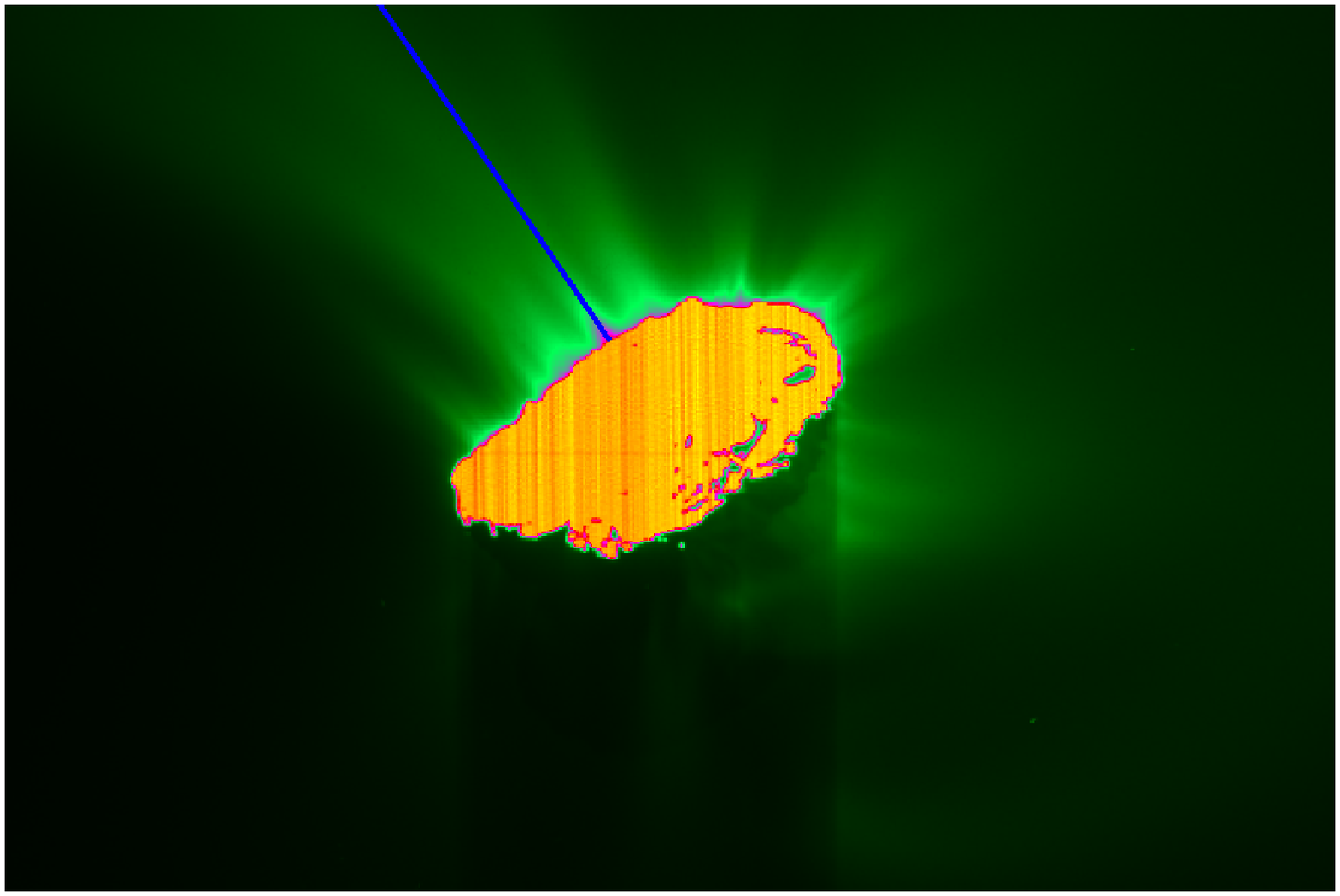}} 
  \subfloat[2015-05-30T16:43:38]{\label{16.43}\includegraphics[scale=0.25]{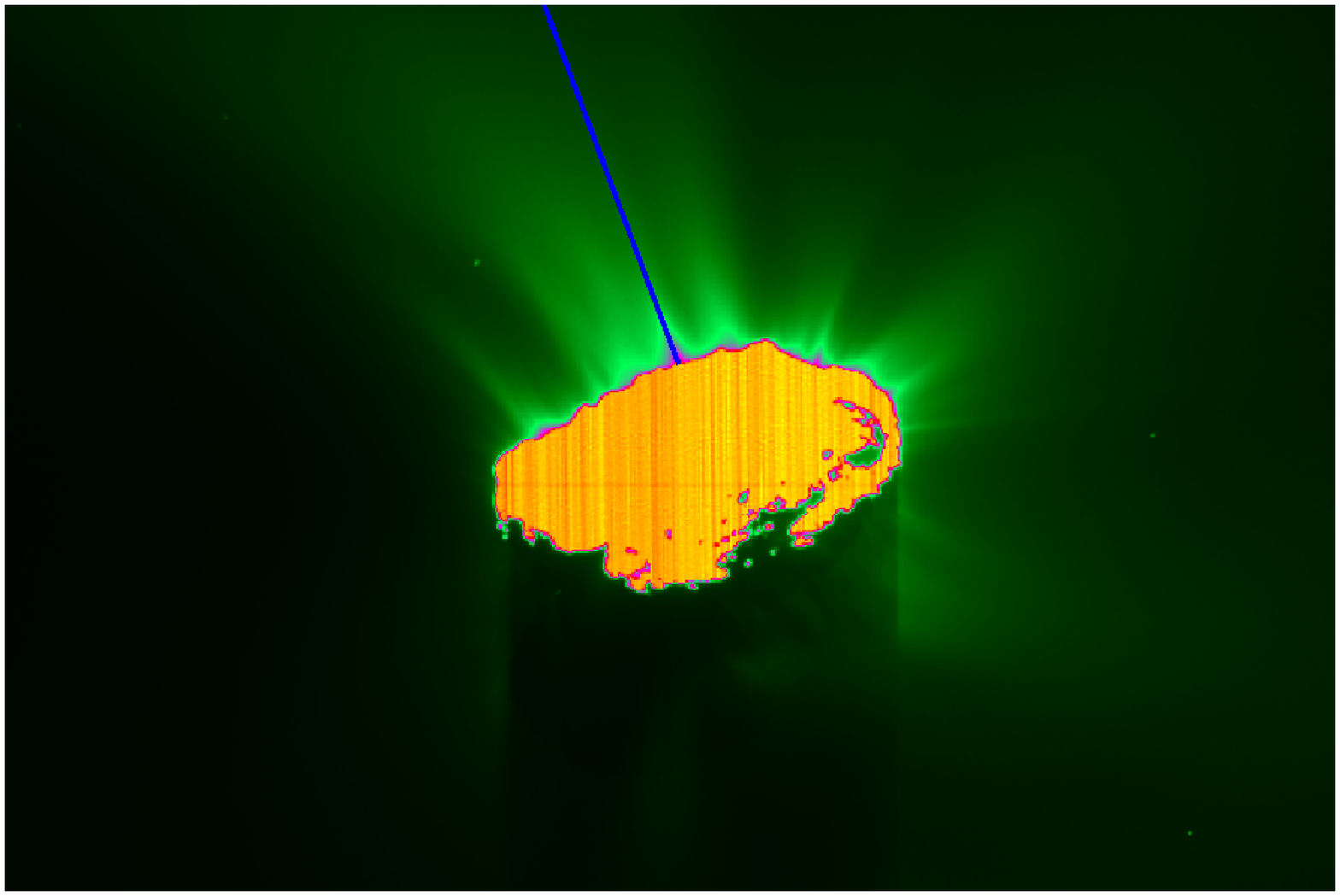}} 
  
	\subfloat[2015-05-30T17:13:38]{\label{17.13}\includegraphics[scale=0.25]{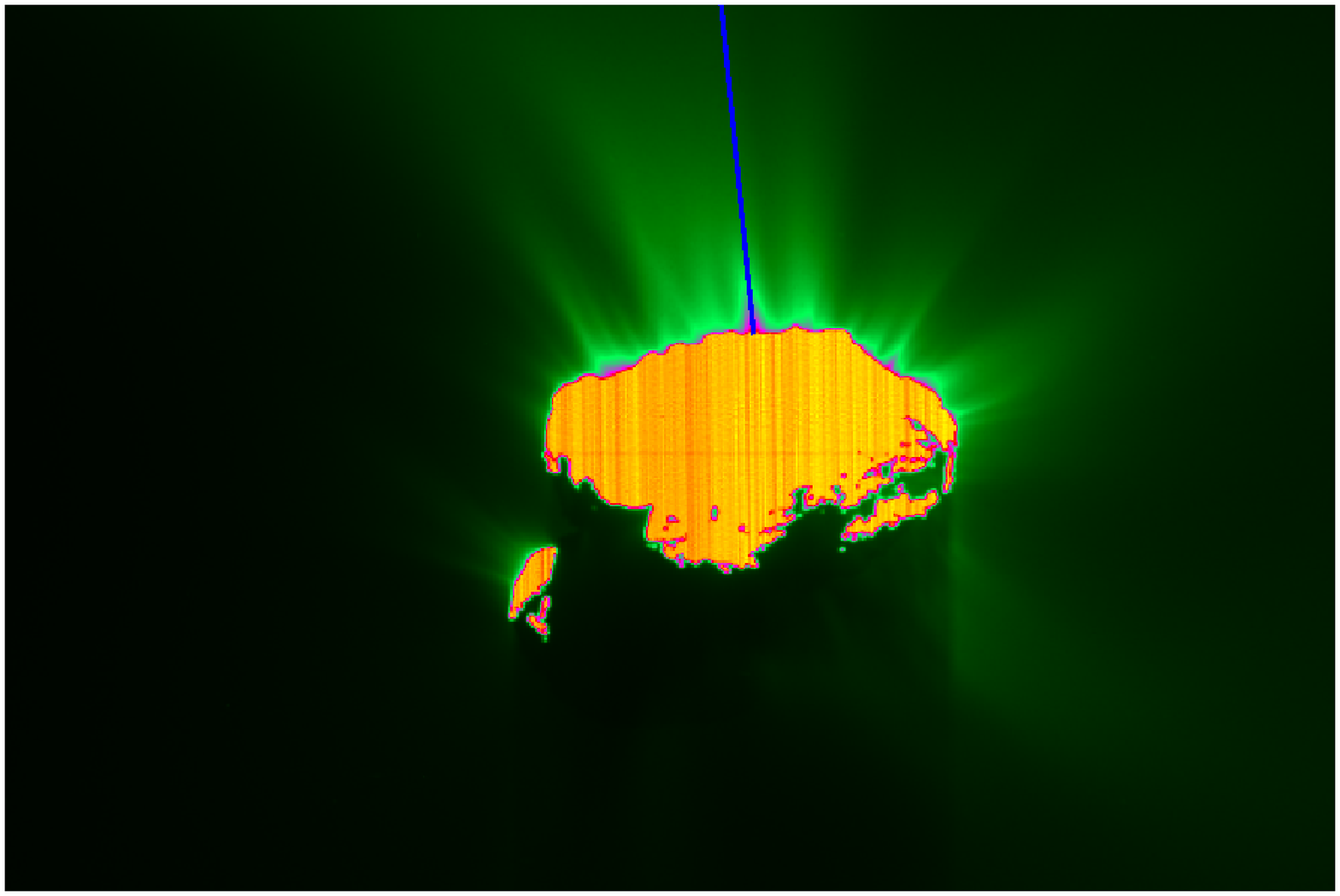}}
  \subfloat[2015-05-30T17:43:38]{\label{17.43}\includegraphics[scale=0.25]{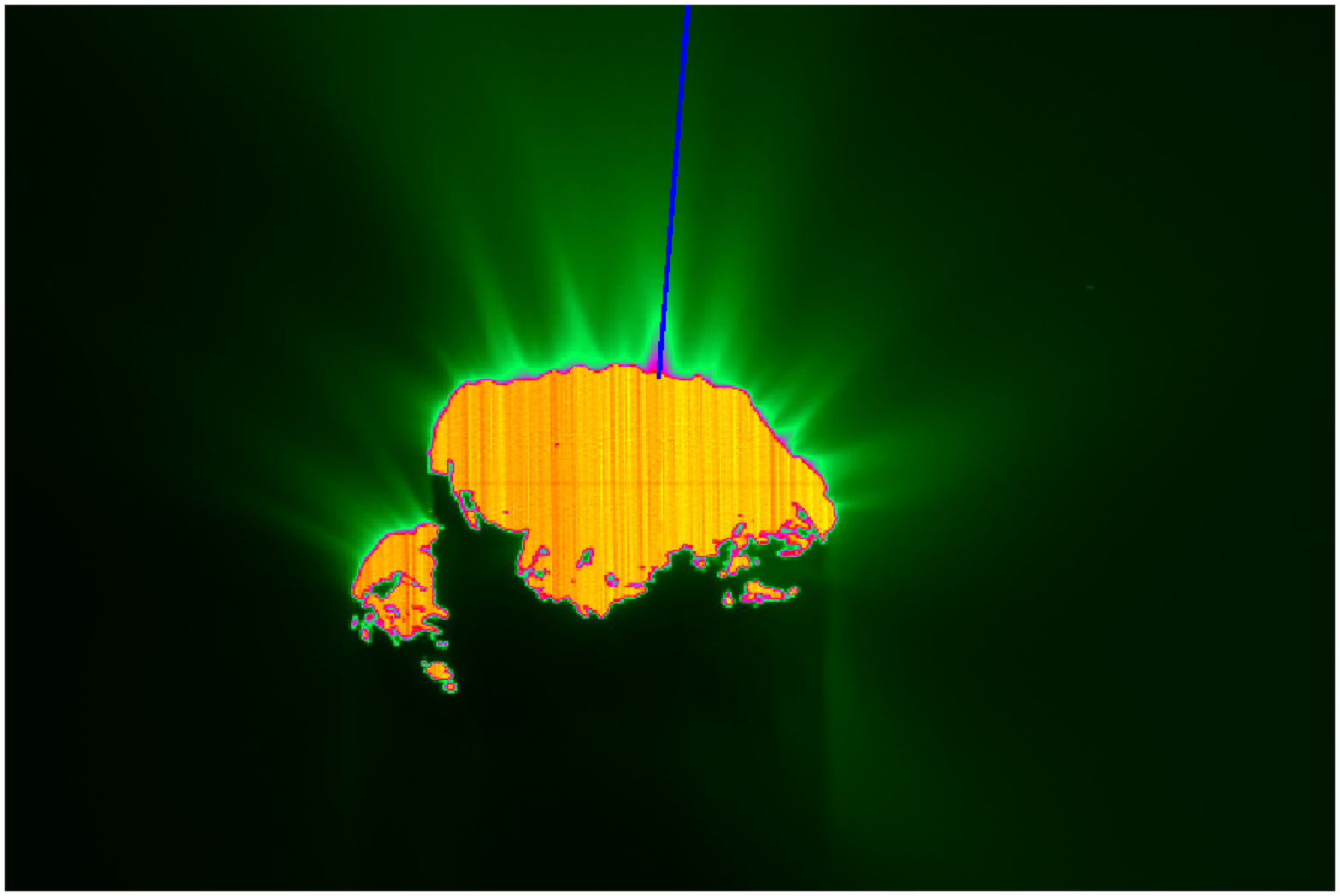}} 
	\subfloat[2015-05-30T18:13:39]{\label{18.13}\includegraphics[scale=0.25]{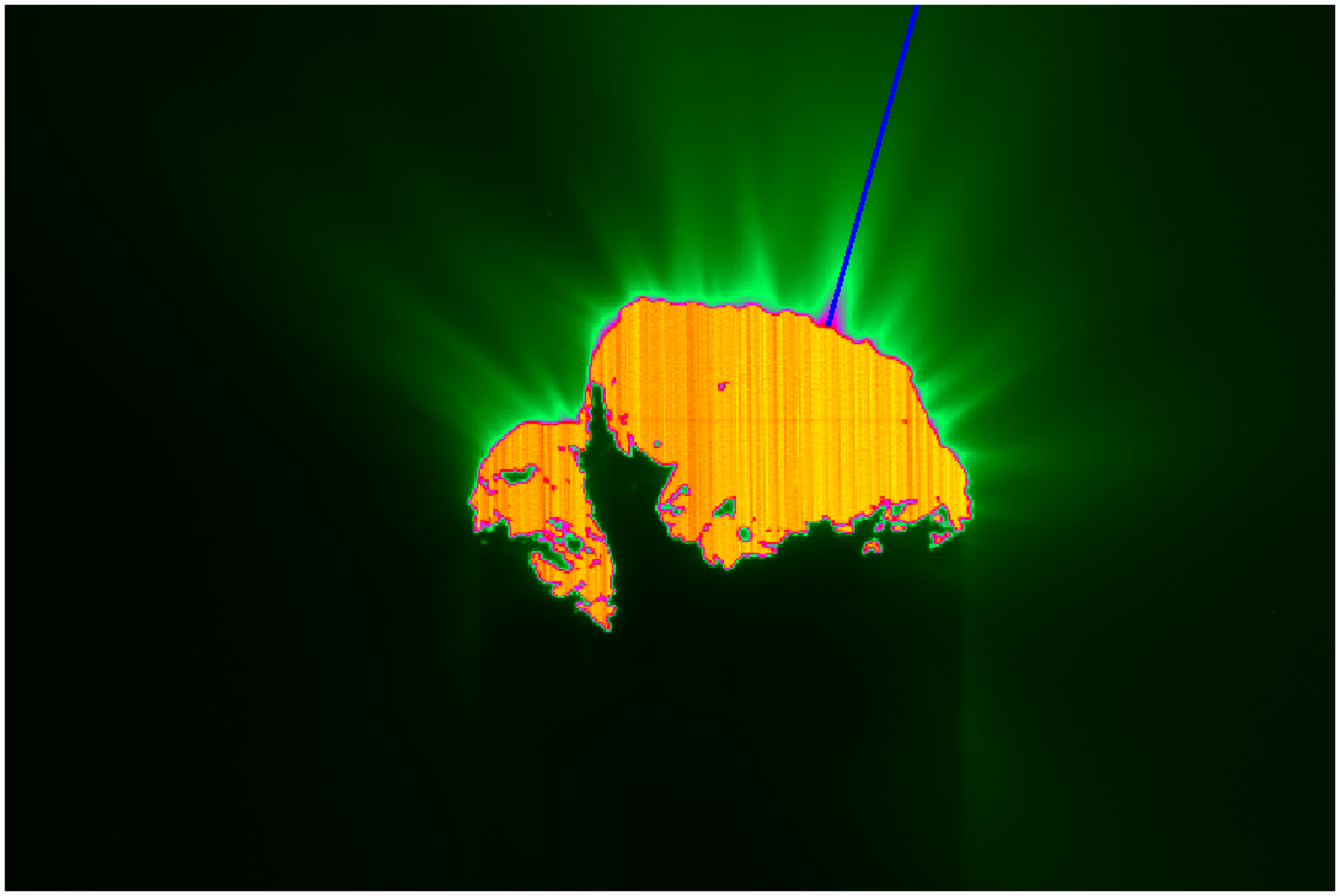}} 
	
	\subfloat[2015-05-30T18:43:38]{\label{18.43}\includegraphics[scale=0.25]{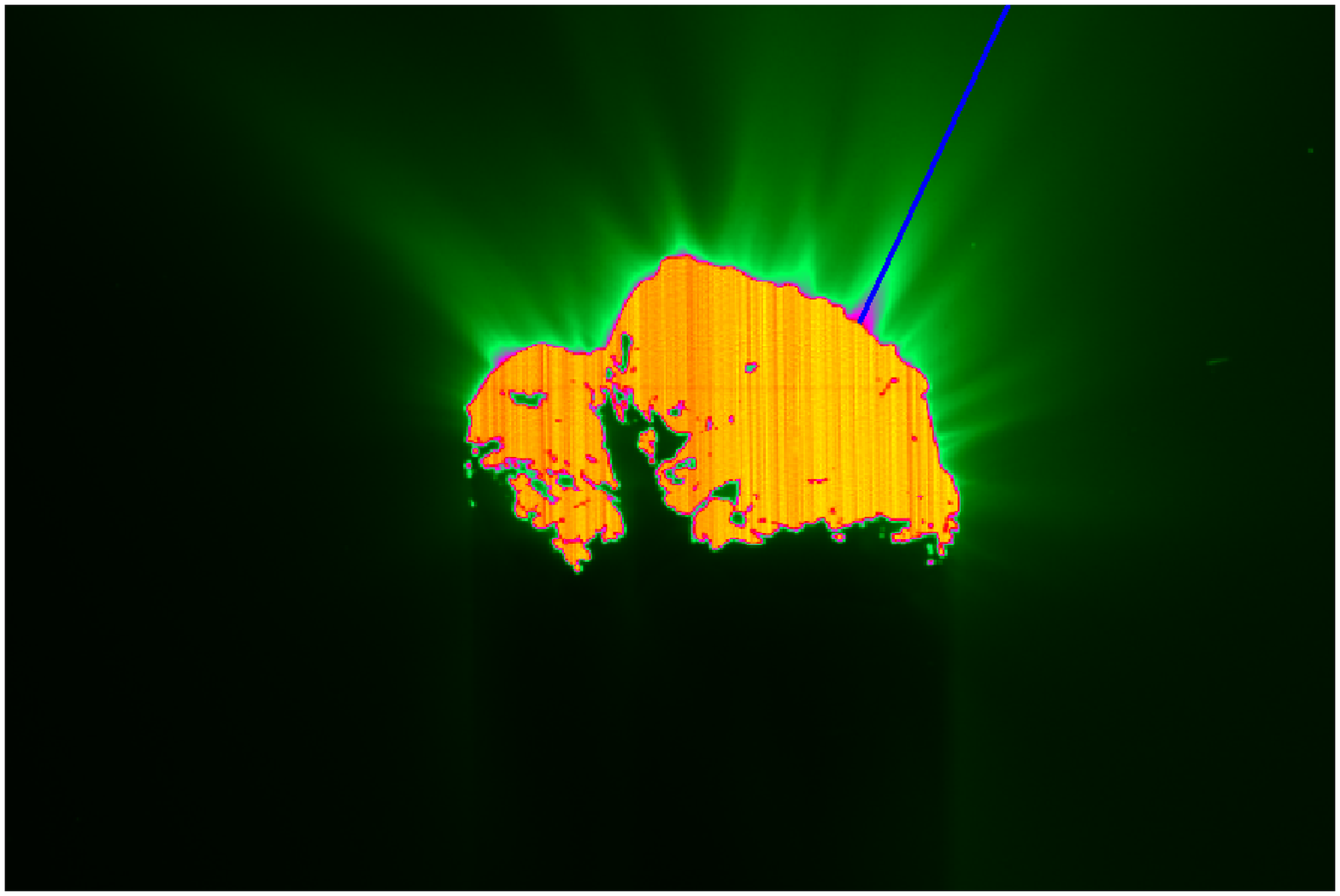}} 
  \subfloat[2015-05-31T04:16:49]{\label{04.16}\includegraphics[scale=0.25]{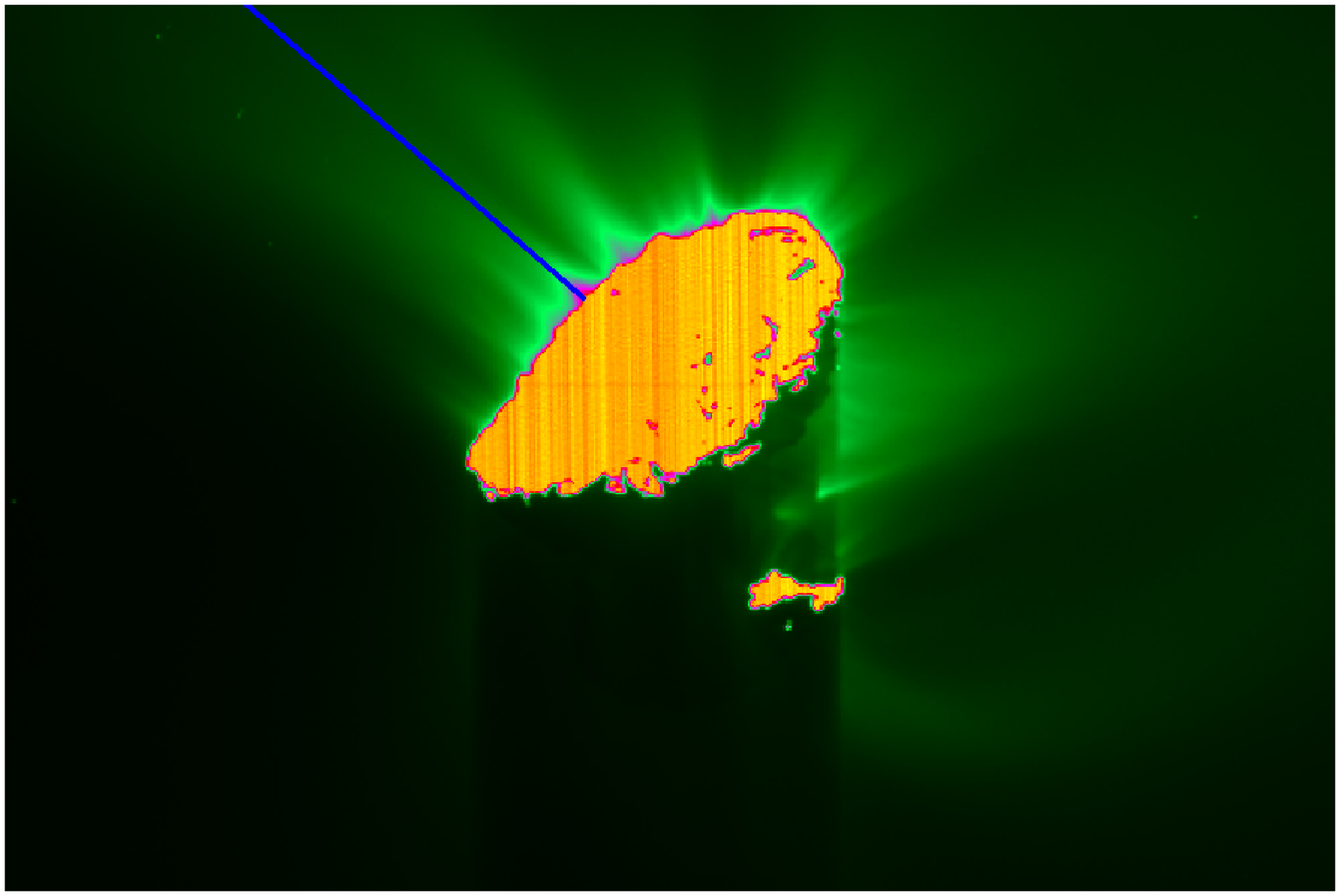}} 
  \subfloat[2015-05-31T04:46:49]{\label{04.46}\includegraphics[scale=0.25]{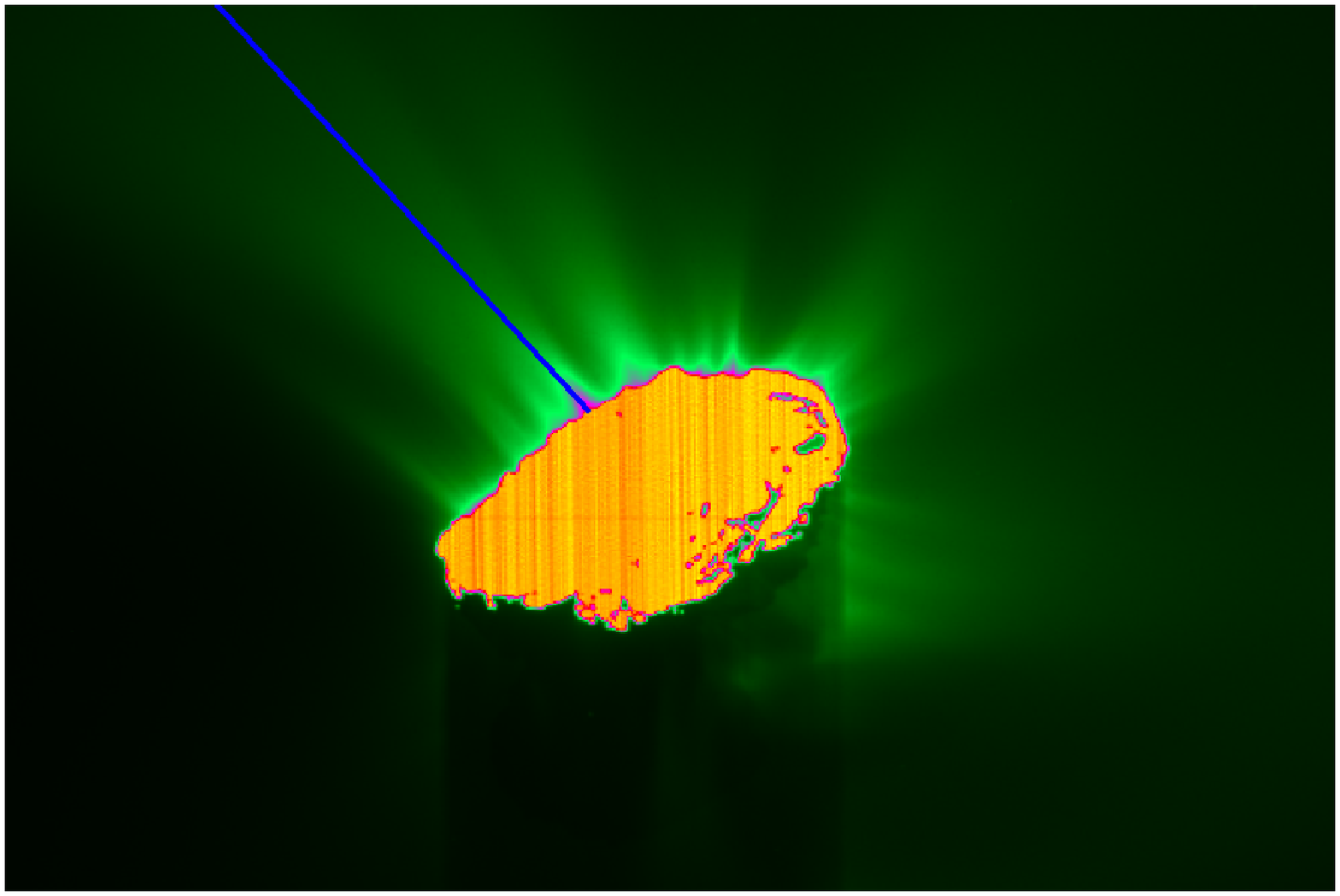}} 
  
	\subfloat[2015-05-31T06:58:22]{\label{06.58}\includegraphics[scale=0.25]{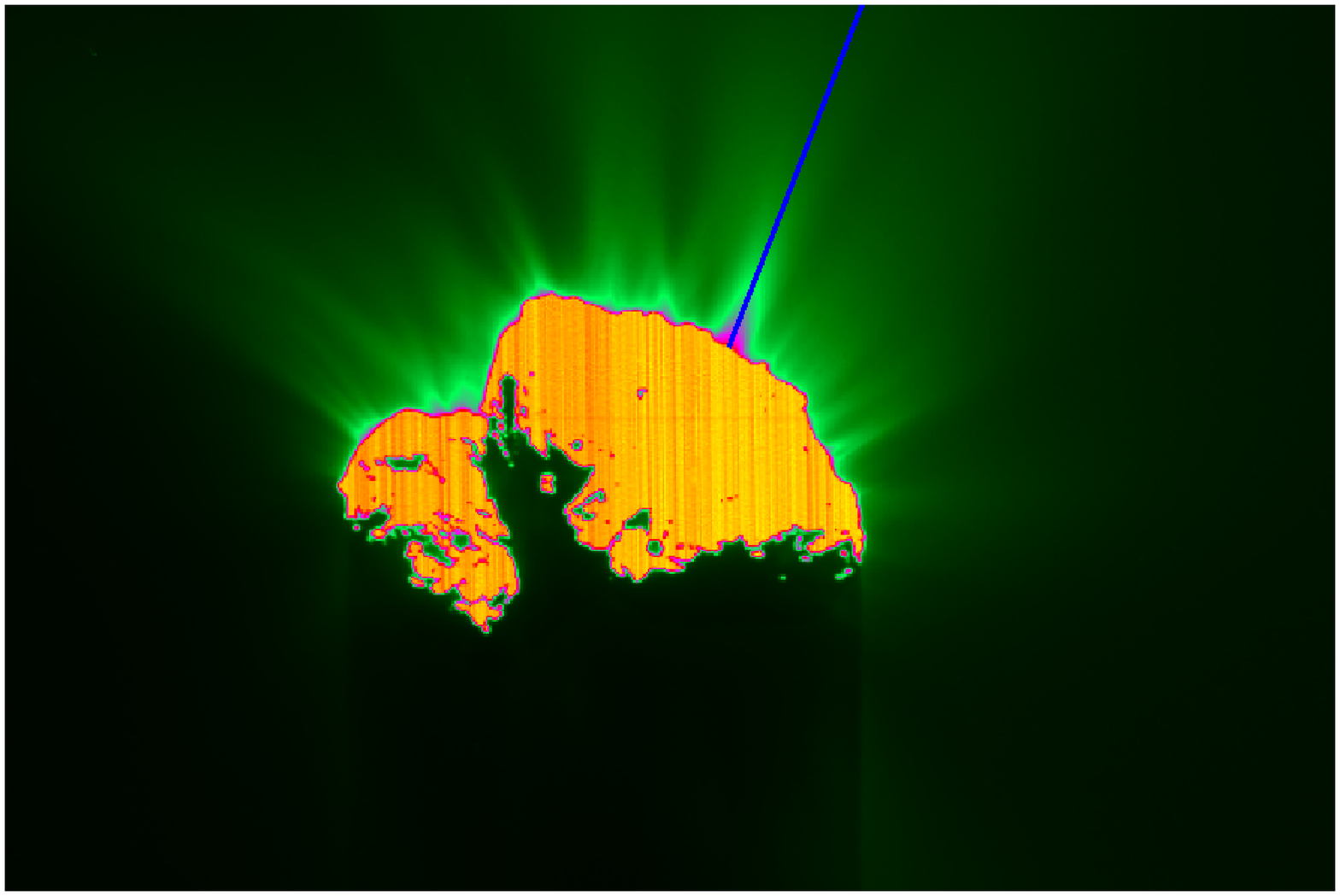}}
  \subfloat[2015-05-31T07:28:22]{\label{07.28}\includegraphics[scale=0.25]{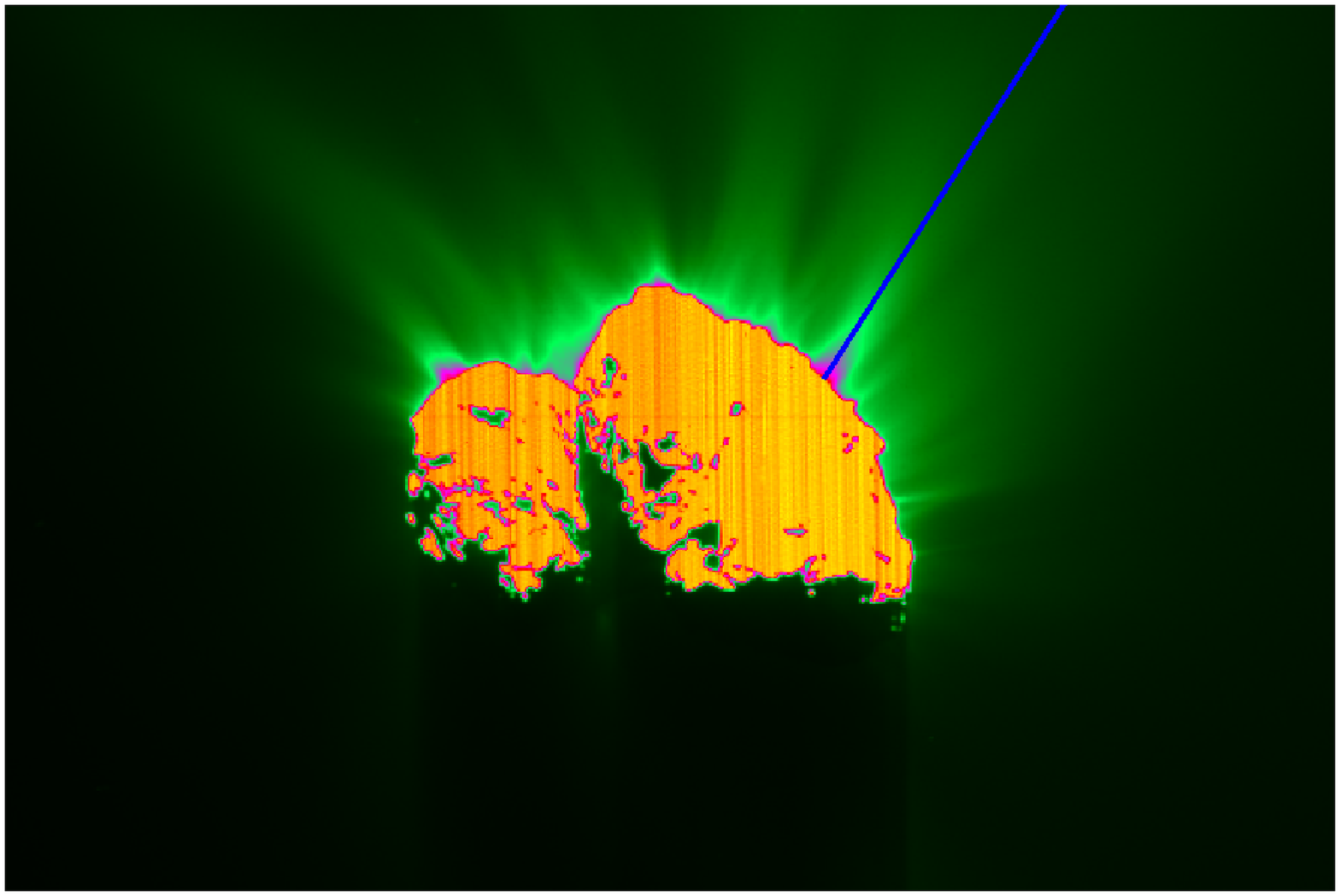}} 

\caption{The OSIRIS 'Dust Monitoring Campaign' - The radial profile
for the jet (blue) and the radial profile for the coma background (green).}
  \label{Image_Jet_Coma}
\end{figure*}

We began by analyzing data from a sequence covering 16h, from UT 15:43:38, May 30th 2015 until UT 07:28:22, May 31st 2015. We exclude data taken during the night time (darkness) at the source of the jet. Only images taken when the big lobe is illuminated, i.e., when the parts of Imhotep that are in the southern hemisphere were illuminated, were used (11 images; Figure \ref{Image_Jet_Coma}). On May 30, 2015 the time interval was 30 min. We did not consider images between UT 20:55:13 May 30th 2015 and UT 3:46:49 May 31st 2015 because the base of the jet was in darkness. On May 31, 2015 the time interval between images was 30 min for the images between UT 04:16:49 - 04:46:49 and UT 06:58:22 - UT 07:28:22, separated by a 2h interval without observations. The image IDs and the relevant circumstances are listed in Table \ref{Obs_VISIBLE}. For all images the heliocentric distance is 1.53 AU, and the resolution is 1.01 $\times$ $\rm 10^{-4}$ rad $\rm pixel^{-1}$. For $\Delta_{S/C}$ around 200 km, the WAC field of view is (FOV) = 40 $\times$ 40 km. No binning was used in collecting or downlinking the images. The dust monitoring sequence includes multiple filters, but we use primarily images taken with a narrow-band filter in the red (Visible), designed to minimize any gaseous emission lines (center wavelength = $\lambda$ = 610nm, FWHM = 9.45nm). It provides a better signal to noise than images in a filter for the near-ultraviolet (375 nm, UV). The red continuum filter was usually used to obtain both a short and a long exposure time, of which we use the longer exposure time, 6.2 sec. \\

\begin{table}
\caption{The OSIRIS 'Dust Monitoring Campaign' - WAC Images - Visible filter (F18). Only images when the south hemisphere is illuminated are taken into account. $\Delta_{S/C}$ is the the spacecraft-comet distance and ID is the Images IDs used as a reference in all images displayed in this work. }
\label{Obs_VISIBLE}
\begin{tabular}{llll}
  \hline
 UT Date & UT Time   & $\rm \Delta_{S/C}$ & id  \\
  &  & (km)  &    \\
  \hline
 May 30, 2015 &15:43:38  & 215.2 & a \\
							&16:13:39   & 214.7 & b \\
							&16:43:38   & 214.3 & c \\
							&17:13:38   & 213.9 & d \\
							&17:43:38   & 213.5 & e \\
							&18:13:39   & 213.1 & f \\
							&18:43:38   & 212.7 & g \\
 May 31, 2015 &04:16:49   & 206.2 & h \\
							&04:46:49   & 206.0 & i \\
							&06:58:22   & 204.9 & j \\
							&07:28:22   & 204.6 & k \\

\hline
\end{tabular}
\end{table}

Figure \ref{Image_Jet_Coma} shows all of the WAC images listed in Table \ref{Obs_VISIBLE}, and the image contrast level in each is adjusted to show the dust jet feature most clearly. In this particular study, we focus only on one jet in the southern part of the Imhotep region (in blue). The first image (\ref{Image_Jet_Coma}a) shows the big lobe and the southern hemisphere illuminated for the first time and we see the apparition of the jet. The jet in blue follows the rotation of the nucleus when the big lobe is illuminated (images \ref{Image_Jet_Coma}a-g). Afterward, the big lobe is on the night side, and we cannot observe the jet. Then the process reverses, and we detect it once again (images \ref{Image_Jet_Coma}h-k). The jet repeats on successive rotations. The images \ref{Image_Jet_Coma}h, \ref{Image_Jet_Coma}i and \ref{Image_Jet_Coma}j were taken after about one full rotation of the nucleus after the images \ref{Image_Jet_Coma}a, \ref{Image_Jet_Coma}b and \ref{Image_Jet_Coma}g, respectively. \\

The rotation of the jet with the nucleus is likely due to the fact that the jet comes from a very specific area of the nucleus, which might have a different composition or morphology than the rest of the nucleus \citep{Lara_2015, Lin_2015, Lin_2016, Vincent_2016}. Importantly, observations with the Visible and Infrared Thermal Imaging Spectrometer (VIRTIS) instrument on the neck region (Hapi) indicate a cyclic sublimation condensation process that follows local illumination conditions during each comet rotation \citep{DeSanctis_2015}. The Microwave Instrument for the Rosetta Orbiter (MIRO) instrument observed the southern hemisphere when it was still in the darkness, during the period August-October 2014 \citep{Choukroun_2015} and September 2014 \citep{Schloerb_2015}. Due to the large difference in the brightness temperatures measured by the two MIRO continuum channels, they hypothesized that ice(s) might have been formed upon cooling after the previous perihelion passage of 67P. They concluded that these regions may be enriched in ice(s) on the surface or within the first tens of centimeters below the surface. This conclusion is in good agreement with the analysis done by \cite{Shi_2015} with the OSIRIS data. The sublimation of water occurs mainly from the illuminated uppermost surface layers, followed by the re-condensation of the water ice when the surface goes into shadow. However, the diurnal re-condensation might be a frost that disappears entirely within an hour after sunrise. The enrichment might be due to the fallback of icy grains mixed in with the dust. The same behavior has been found for comet 9P/Tempel 1 \citep{Vincent_2010,Farnham_2013,Feaga_2007} and comet 103P/Hartley 2 \citep{Mueller_2013}. In this manner, the surface layers become enriched in water ice and become more energetic when the comet is closer to the Sun.  \\

\subsection{Radial profiles}
\label{Rad_Pro}
\begin{figure}
\centering
\includegraphics[scale=0.38]{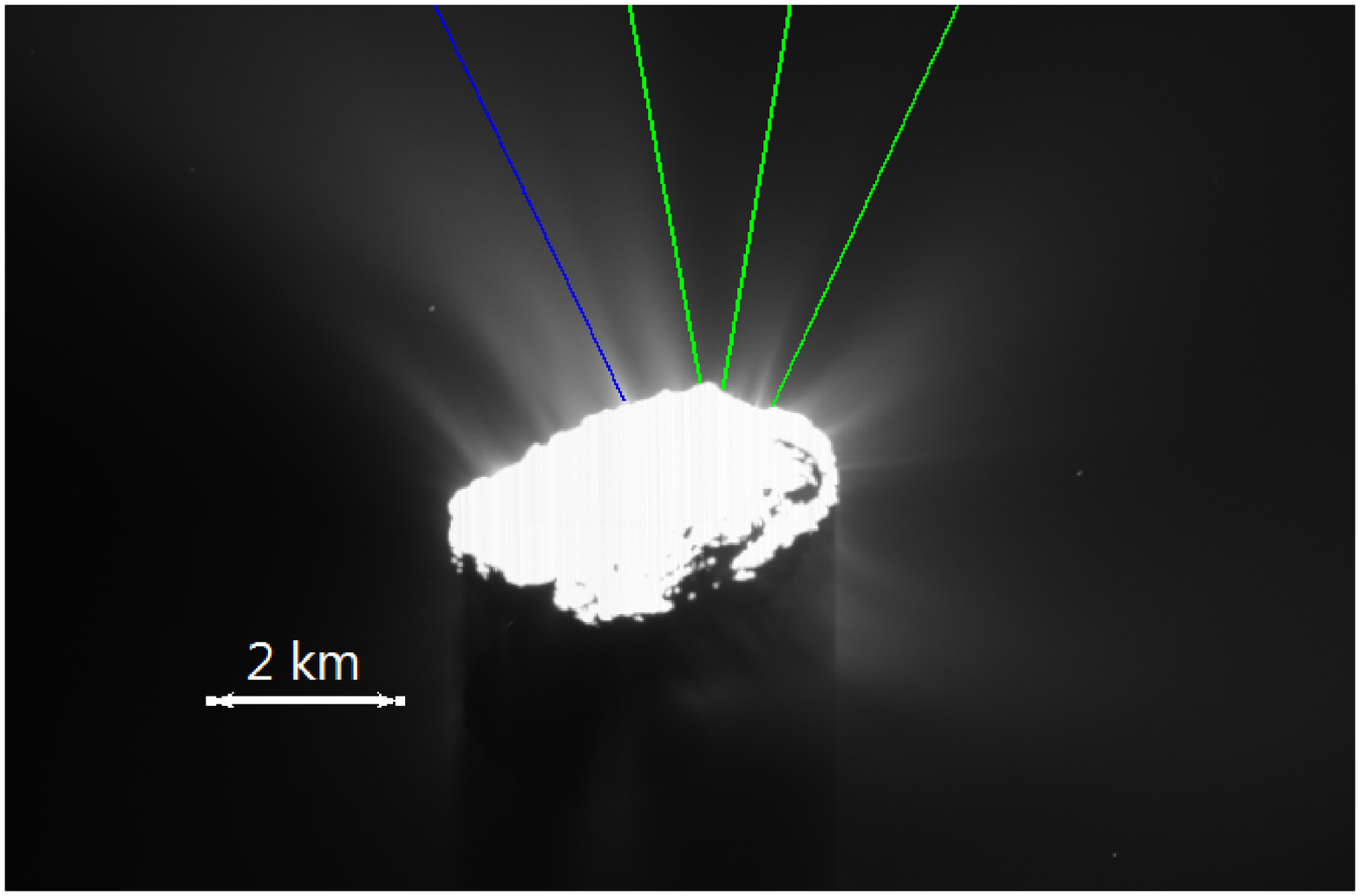}
  \vspace{0.3in}
\caption{2015-05-30T16:43:38 - In blue is the radial profile for the jet and in green are the radial profiles for the coma background. }
\label{Method}
\end{figure}

In the present article, we aim to study the brightness distribution of the dust jet as a function of distance from the nucleus. We assume $B$ $\propto$ $D^{\beta}$, where $B$ is the brightness of the jet, $D$ is the radial distance from the surface of the nucleus and $\beta$ is the dimensionless slope of log$B$ vs. log$D$. Multiple methods were used to obtain the coma background, which must be removed from the profile in the jet. \cite{Lin_2015} computed the azimuthally averaged background brightness distribution of the faint coma surrounding 67P. However, the coma structure is clearly non-isotropic and time-dependent \citep{Lara_2015, Tubiana_2015, Mottola_2014}. The time variability is likely due to the rotation that changes the surface temperature of the comet following the insolation \citep{Capaccioni_2015, DeSanctis_2015} or the geometry of the observations \citep{Sierks_2015}. We average 3 radial profiles of the coma for each image in the same area as the jet, as shown in Figure \ref{Method}. The radial profile is taken from the individual pixels along the center-line of the jet. We assume that the jet originates on the comet surface and end after 10 km \citep{Lin_2015, Tozzi_2004}. The jet brightness was below the background noise after 10 km. We did not consider the radial profile in the inner coma (< 1km from the surface) to avoid possible straylight contamination from the nucleus. The coma background is subtracted from the radial profile of the jet. The same study was made with the UV filter. The radial profile of the jet in the Visible and UV filters are identical, which means that there is no indication of changing size or composition of the particles along the jet.  \\

\begin{figure*}
  \centering
  \subfloat[2015-05-30T15:43:38]{\label{15.43}\includegraphics[scale=0.20]{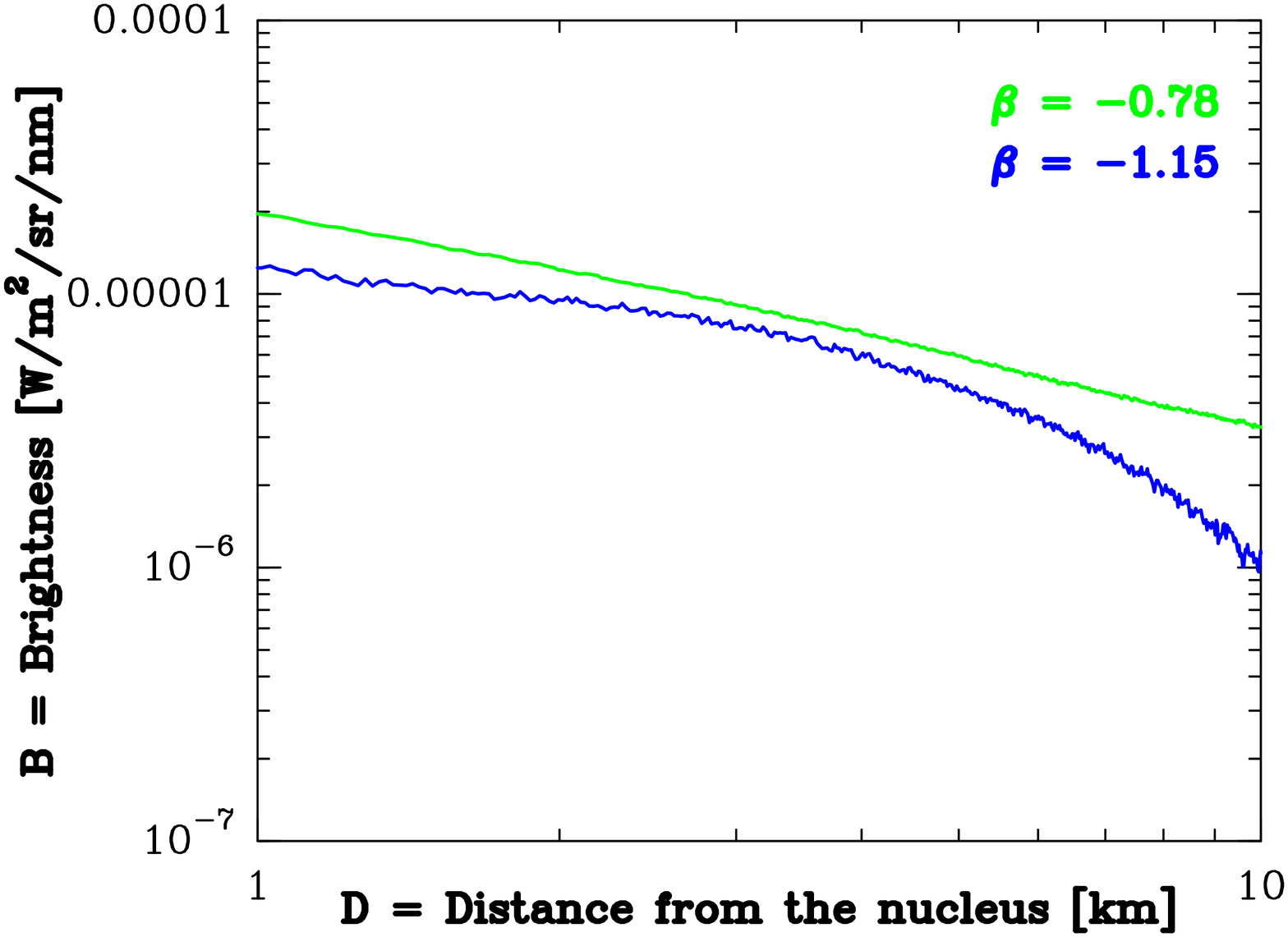}} 
  \subfloat[2015-05-30T16:13:39]{\label{16.13}\includegraphics[scale=0.20]{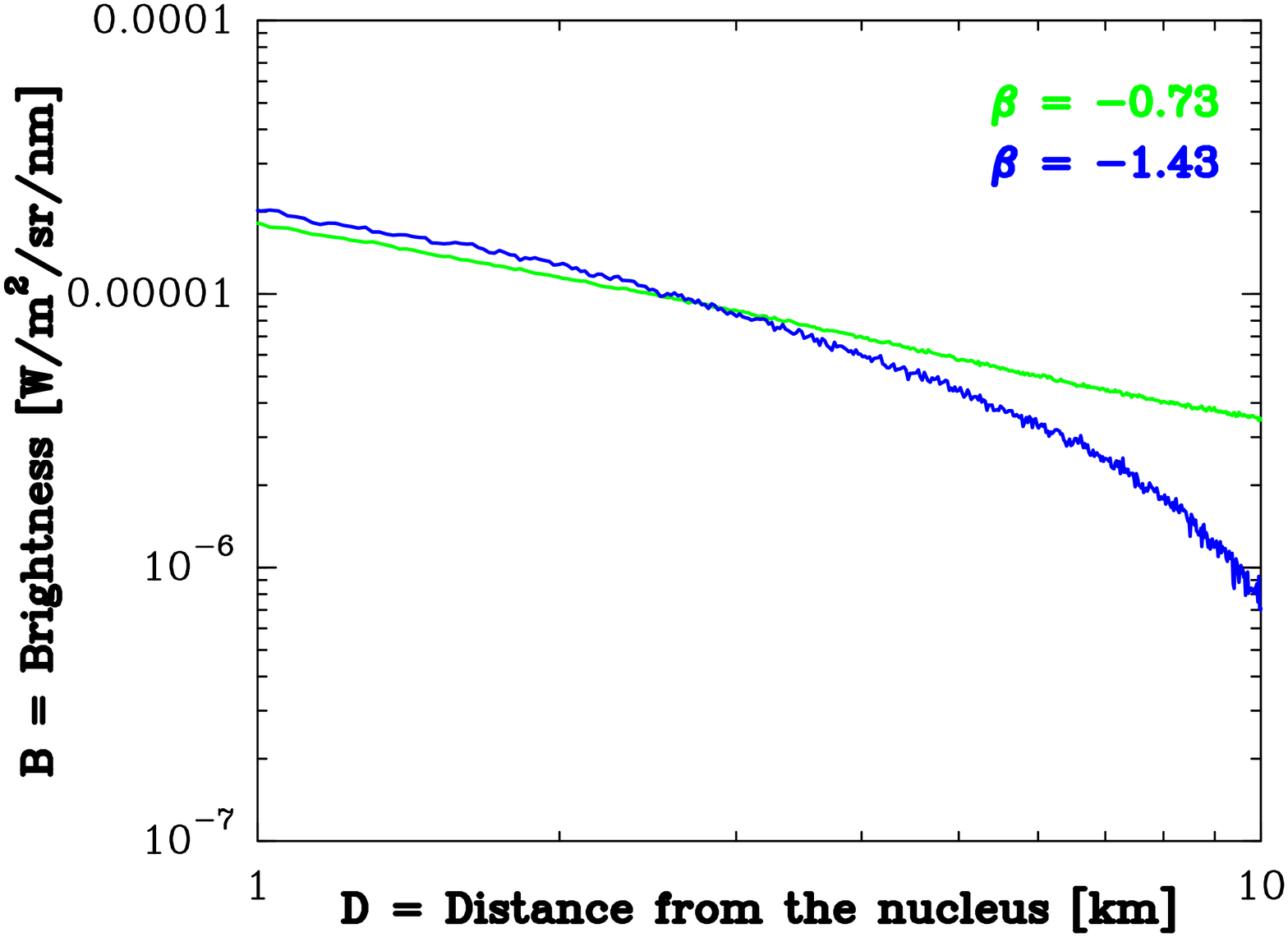}} 
  \subfloat[2015-05-30T16:43:38]{\label{16.43}\includegraphics[scale=0.20]{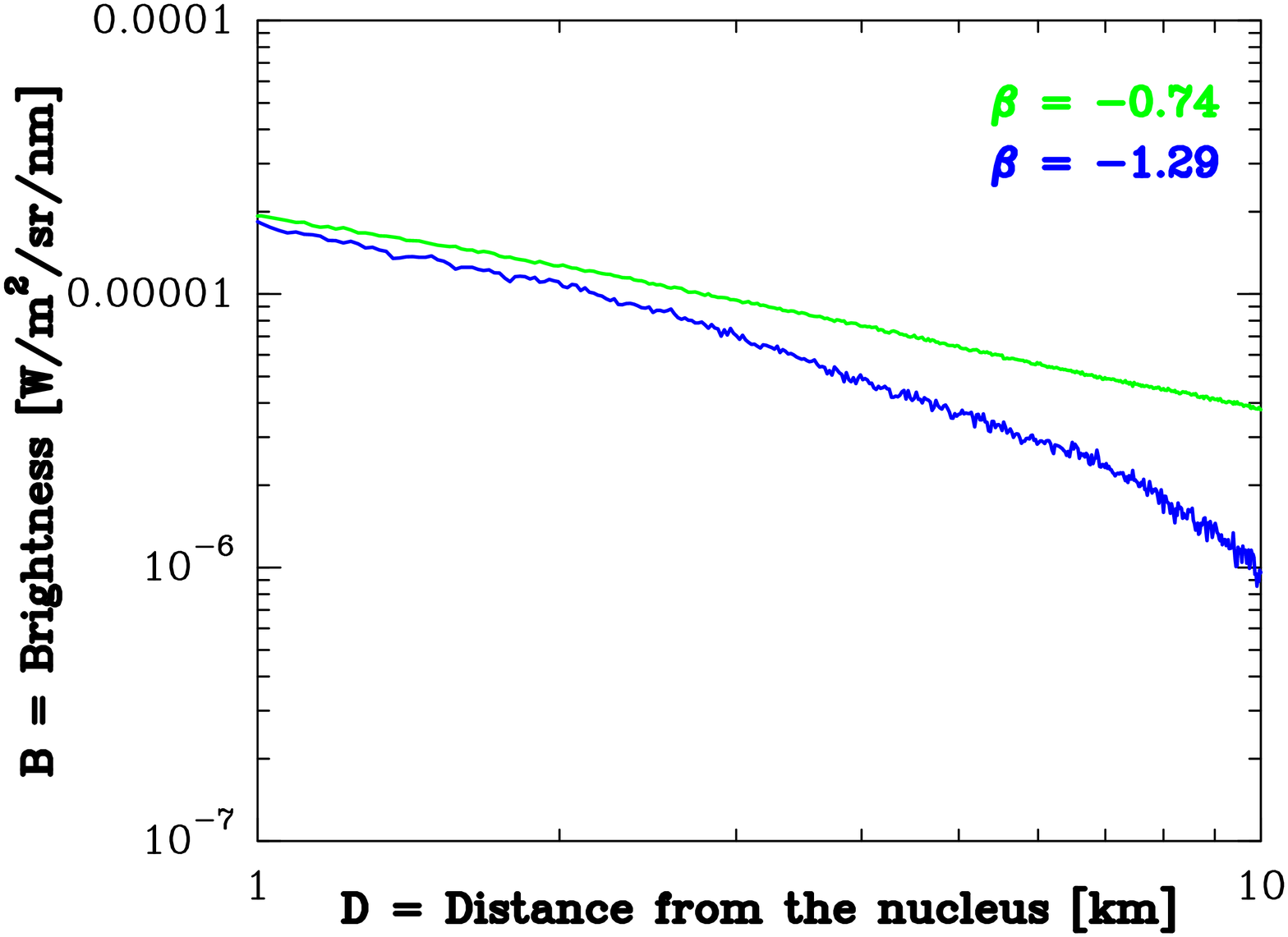}} 
  
	\subfloat[2015-05-30T17:13:38]{\label{17.13}\includegraphics[scale=0.20]{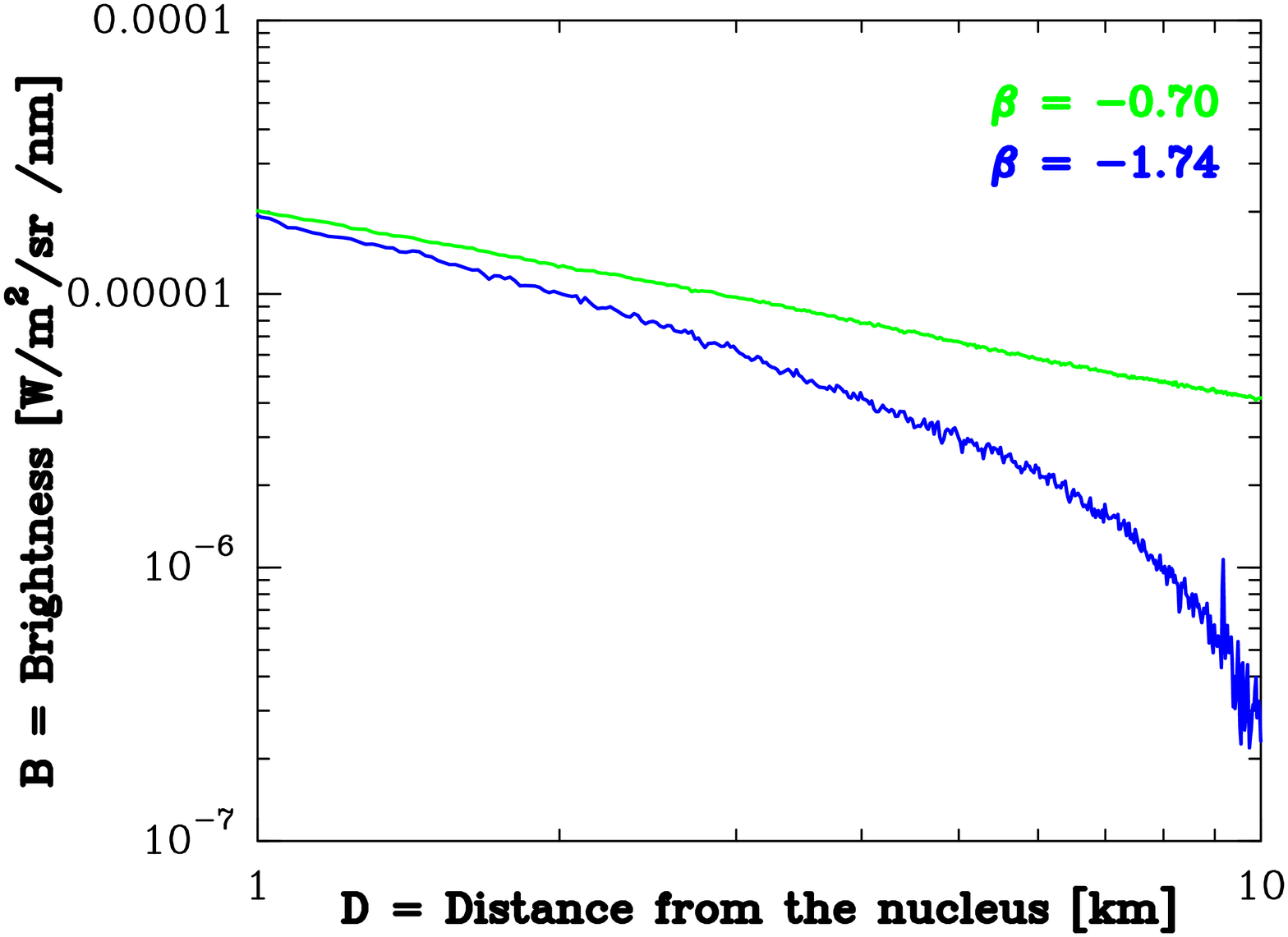}}
  \subfloat[2015-05-30T17:43:38]{\label{17.43}\includegraphics[scale=0.20]{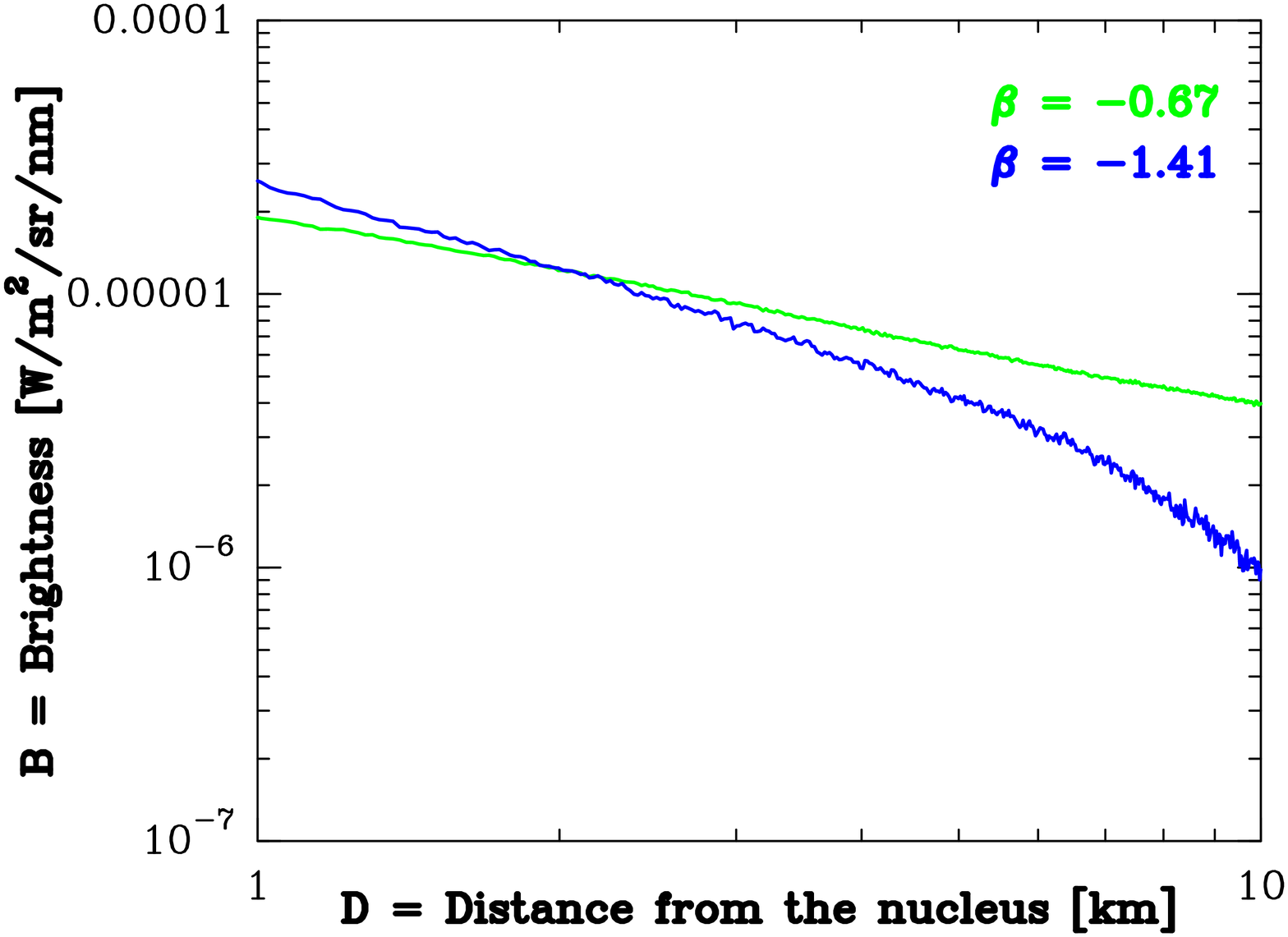}} 
	\subfloat[2015-05-30T18:13:39]{\label{18.13}\includegraphics[scale=0.20]{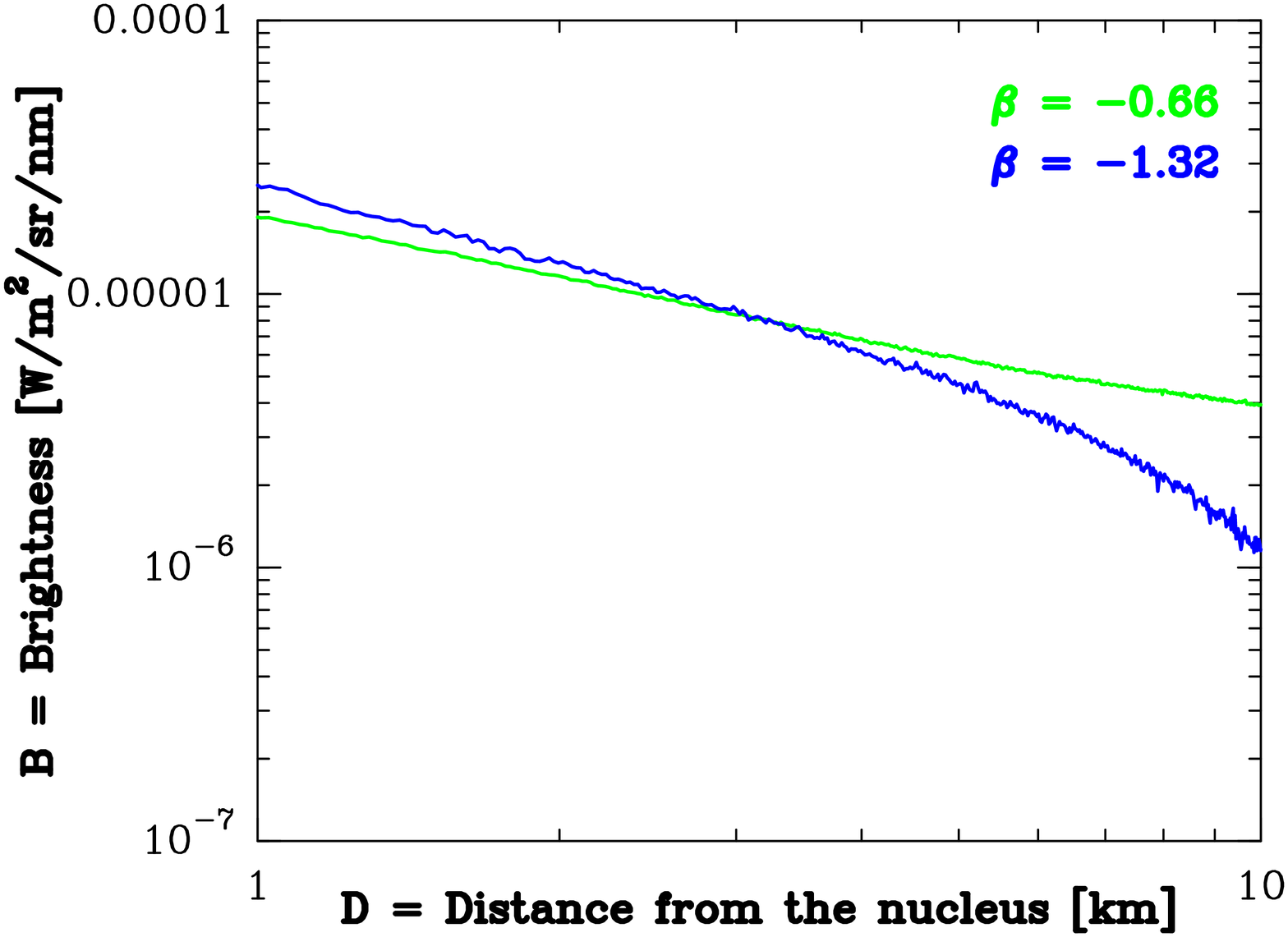}} 
	
	\subfloat[2015-05-30T18:43:38]{\label{18.43}\includegraphics[scale=0.20]{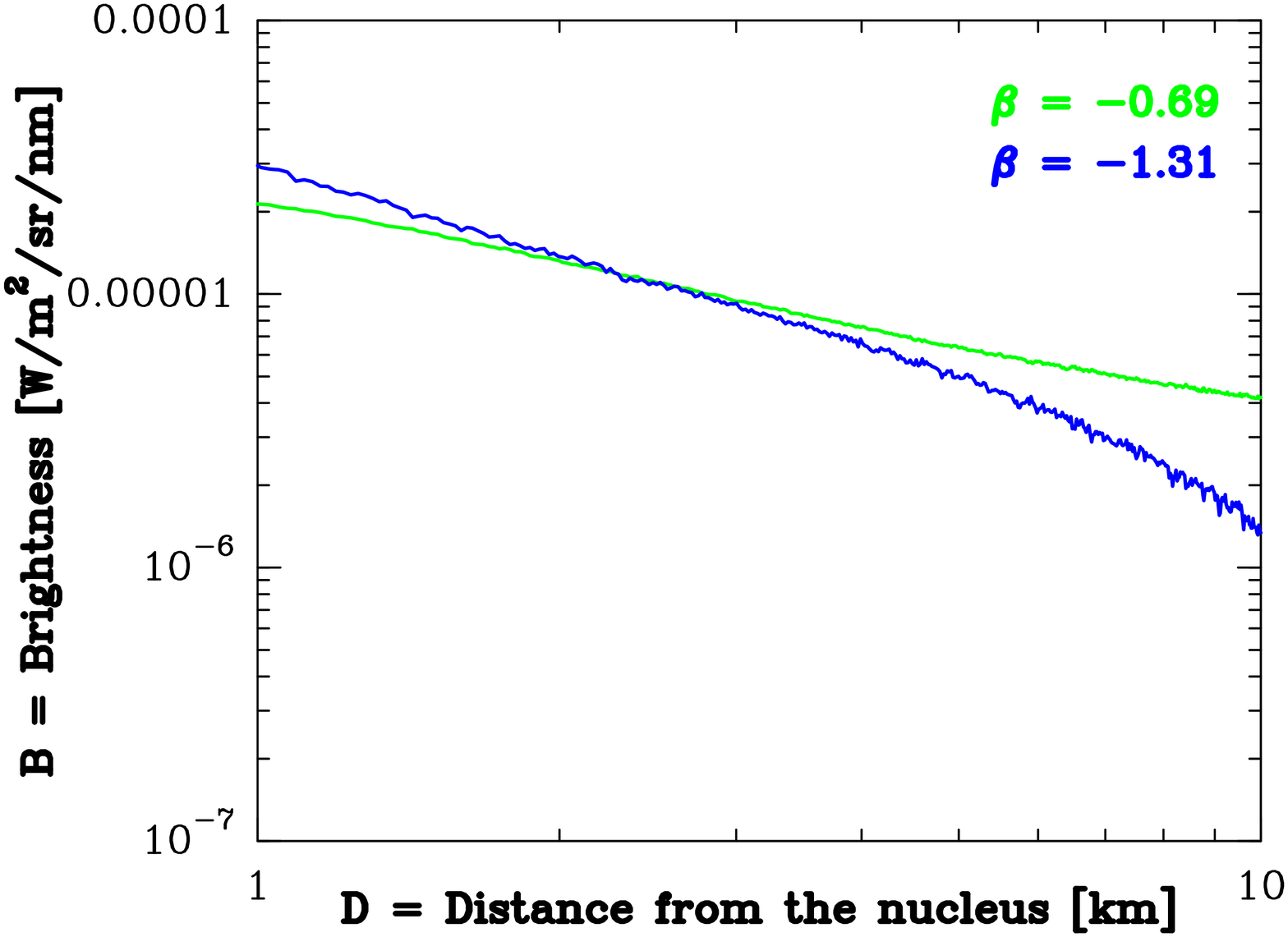}} 
  \subfloat[2015-05-31T04:16:49]{\label{04.16}\includegraphics[scale=0.20]{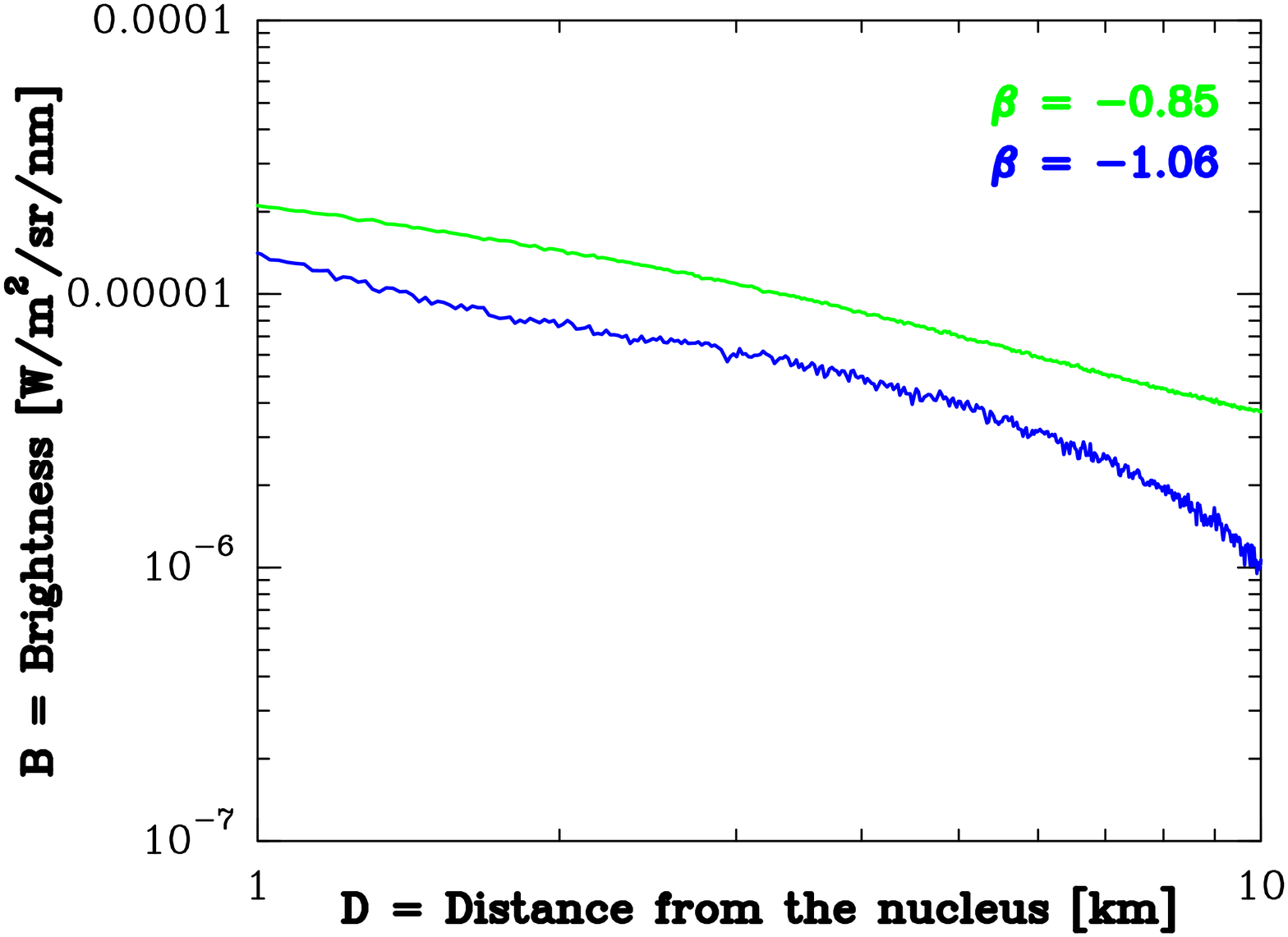}} 
  \subfloat[2015-05-31T04:46:49]{\label{04.46}\includegraphics[scale=0.20]{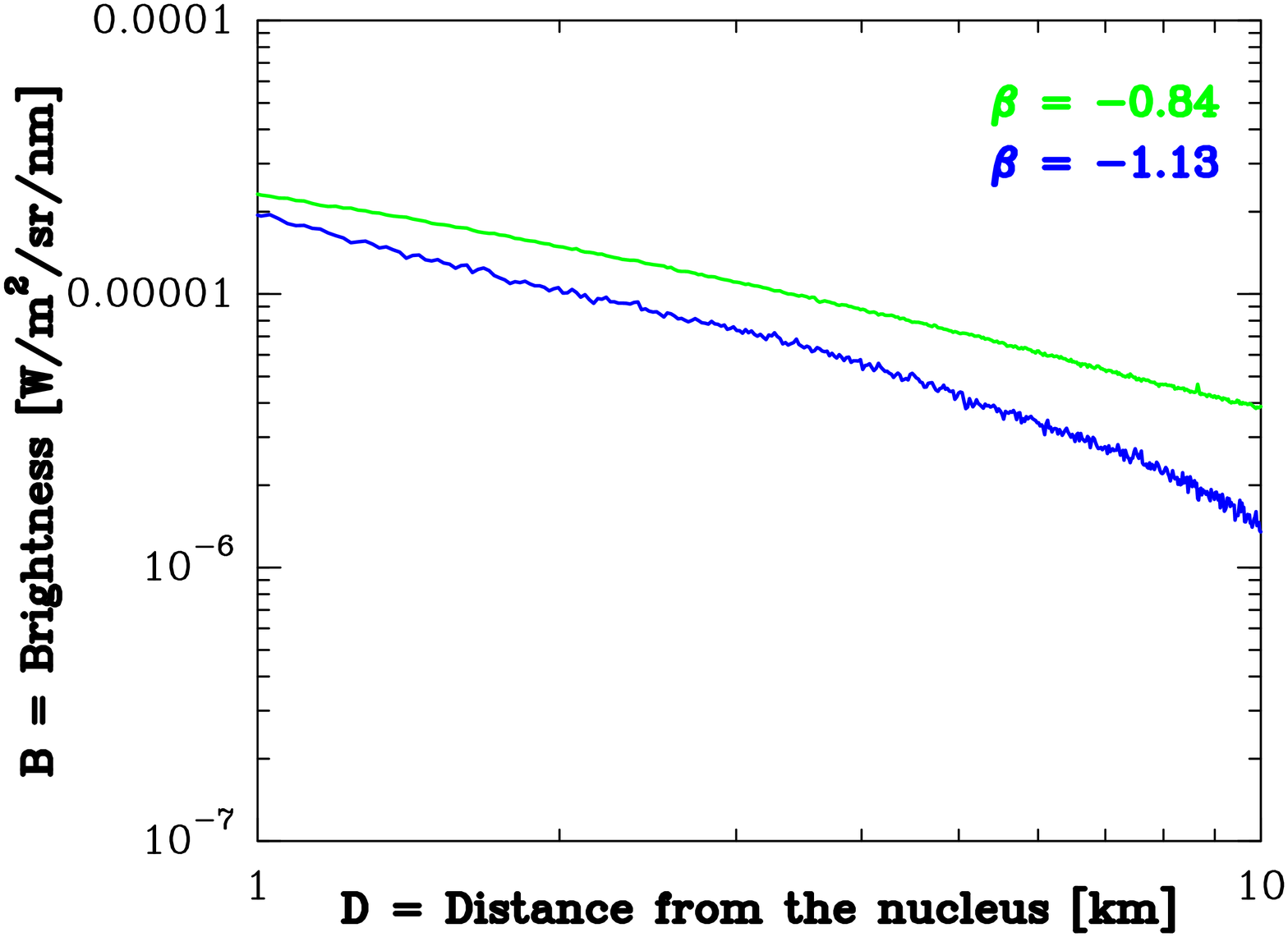}} 
  
	\subfloat[2015-05-31T06:58:22]{\label{06.58}\includegraphics[scale=0.20]{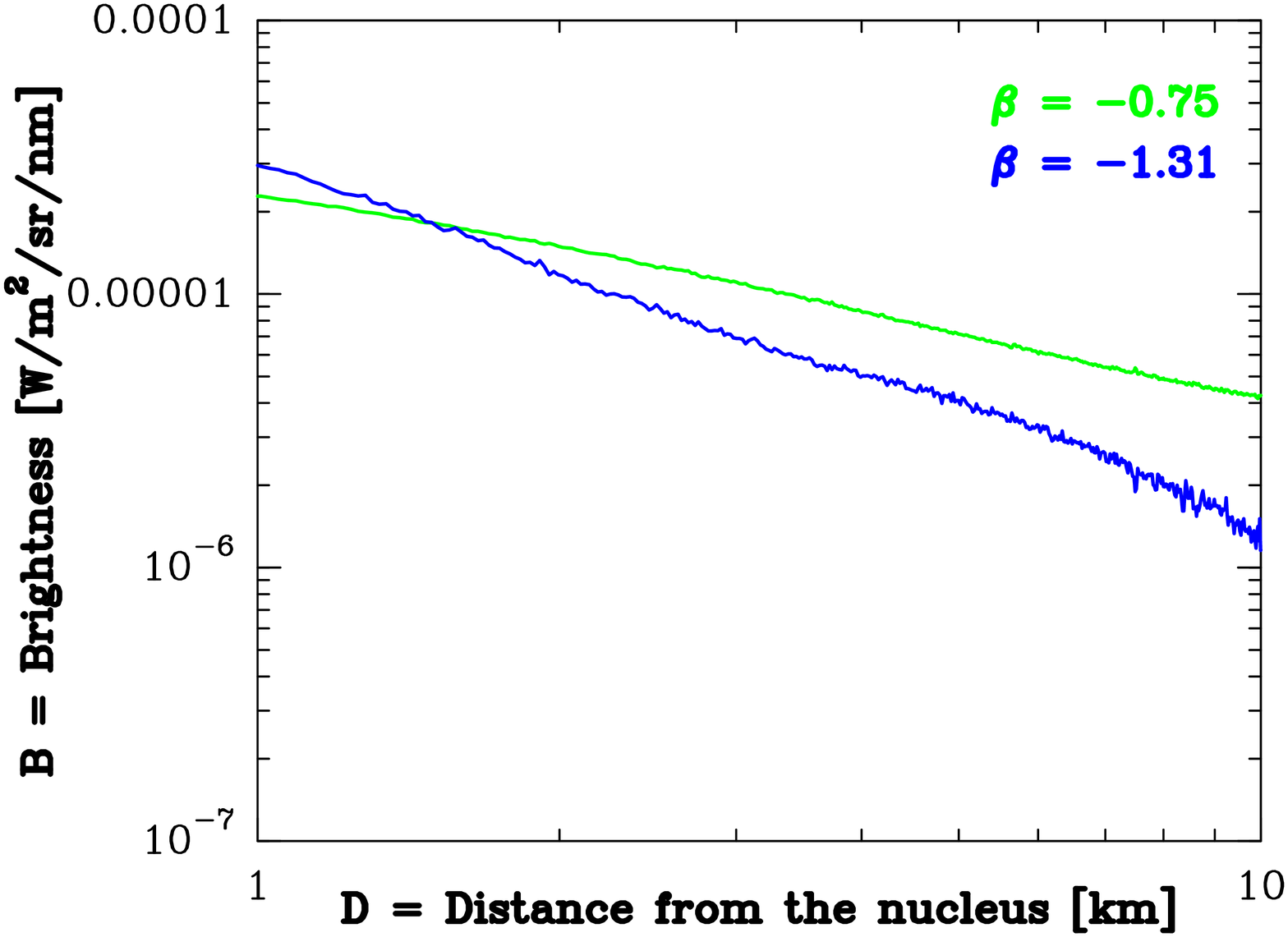}}
  \subfloat[2015-05-31T07:28:22]{\label{07.28}\includegraphics[scale=0.20]{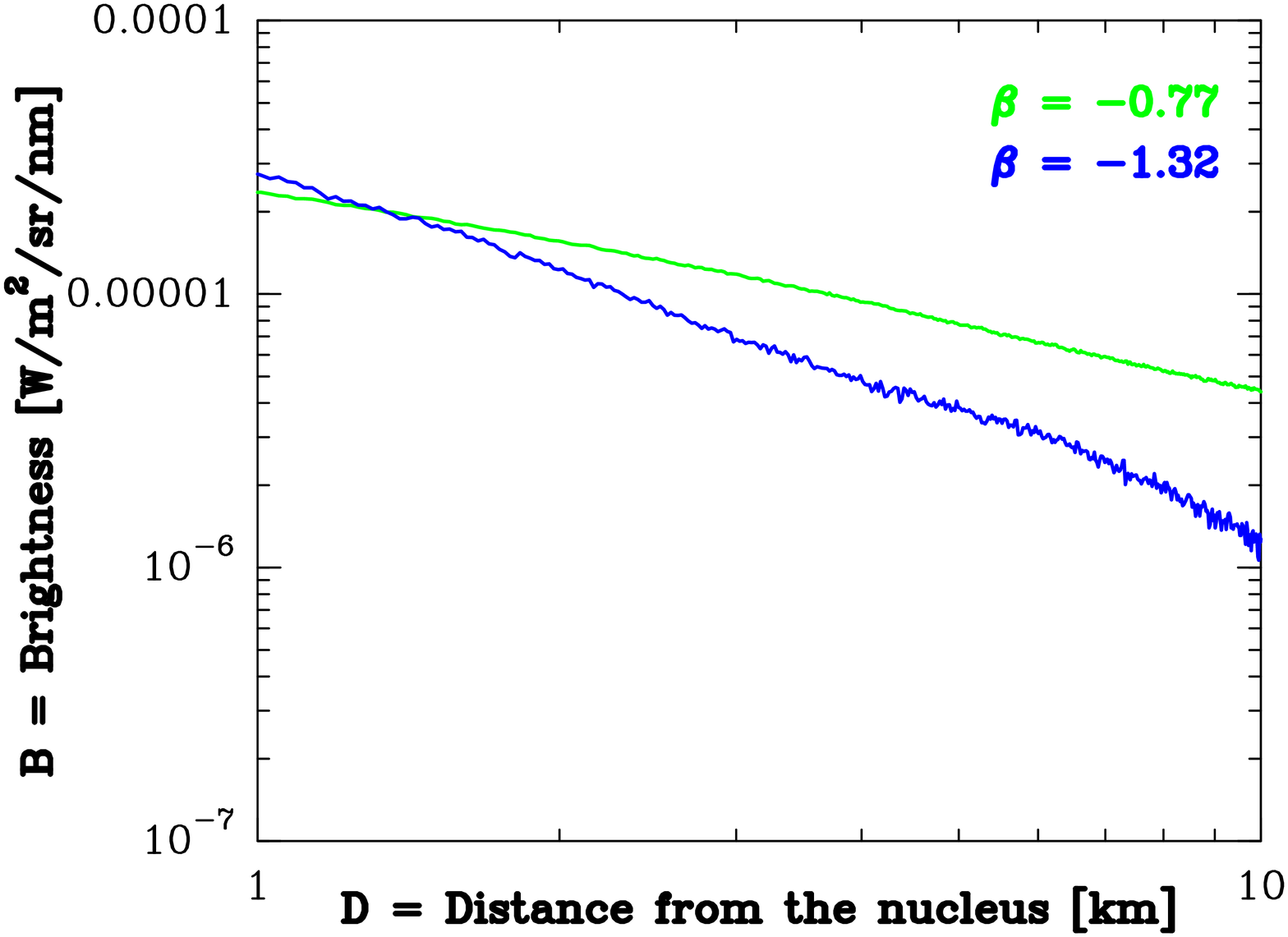}} 

\caption{The OSIRIS 'Dust Monitoring Campaign' - Visible filter. In blue is the radial profile for the jet and in green is the radial profile for the coma background. }
  \label{Vis_Jet_Coma}
\end{figure*}

Figure \ref{Vis_Jet_Coma} shows the radial brightness of the jet (after subtraction of the background coma) and of the background coma for all the WAC images listed in Table \ref{Obs_VISIBLE} and shown in Figure \ref{Image_Jet_Coma}. We find the average brightness slope of the jet to be $\beta$ = -1.31 $\pm$ 0.17, whereas the one corresponding to the background coma is $\beta$ = -0.74 $\pm$ 0.06. On average, the radial profile of the jet is much steeper than the radial brightness of the coma. Also,the brightness profile of the jet is curved (between 4 and 10 km from the surface) in log-log space; whereas, the background coma is linear in log-log space. \\

The MIRO instrument studied the submillimeter continuum (which is proportional to the column density of dust) from 2 August 2015. Observations made between 06:55 and 10:19  \citep{Hofstadter_2016}. \cite{Hofstadter_2016} also found curvature after 6 km from the surface and derived a slope equal to $\beta$ = -1.64, indicating less emission than expected as we move away from the nucleus. \cite{Hofstadter_2016} explained that acceleration due to gas drag including preferential acceleration of the smaller particles and the rapid cooling of the dust within the inner few kilometers seems insufficient, and it cannot explain the data. \cite{Hofstadter_2016} concluded that only the destruction of most large particles within 10 km (by fragmentation or sublimation) can, by itself, fit the observations. This conclusion obtained with MIRO and the value of the slopes are in good agreement with the results we show later in this article. Even if these observations are made closer to the Sun, the MIRO instrument is detecting larger particles, and the comparison between the two instruments has to be checked in more details in the future. \\

\cite{Lin_2015} suggested that the flat brightness slope of the coma (0.41-0.59) from August to September 2014 might be characteristic of the bulk in the inner coma, which consists of large-sized grains with an organic composition \cite{Capaccioni_2015}. \cite{Lin_2015}, also suggested the steeper brightness slope of the jet (0.95-1.48) might be due to sublimation or fragmentation or to the time variability in the gas flow due to the insolation condition. In the past studies from ground-based observation of 67P, a brightness slopes of the background coma around 1.5 had been deduced \citep{Lara_2011,Hadamcik_2010}. A problem with comparing ground-based observation and high resolution images from the OSIRIS cameras was the size of the coma. From the DIXI mission measurements on comet 103P/Hartley 2, \cite{Protopapa_2014} reported a slope close to 1 for the coma background and > 4.2 for the jets, which they interpreted as evidence of sublimation of micron-sized water ice grains.\\

\subsection{Source location of the jet}
\label{Jet_Location}
\begin{figure}
\centering
\includegraphics[scale=0.60]{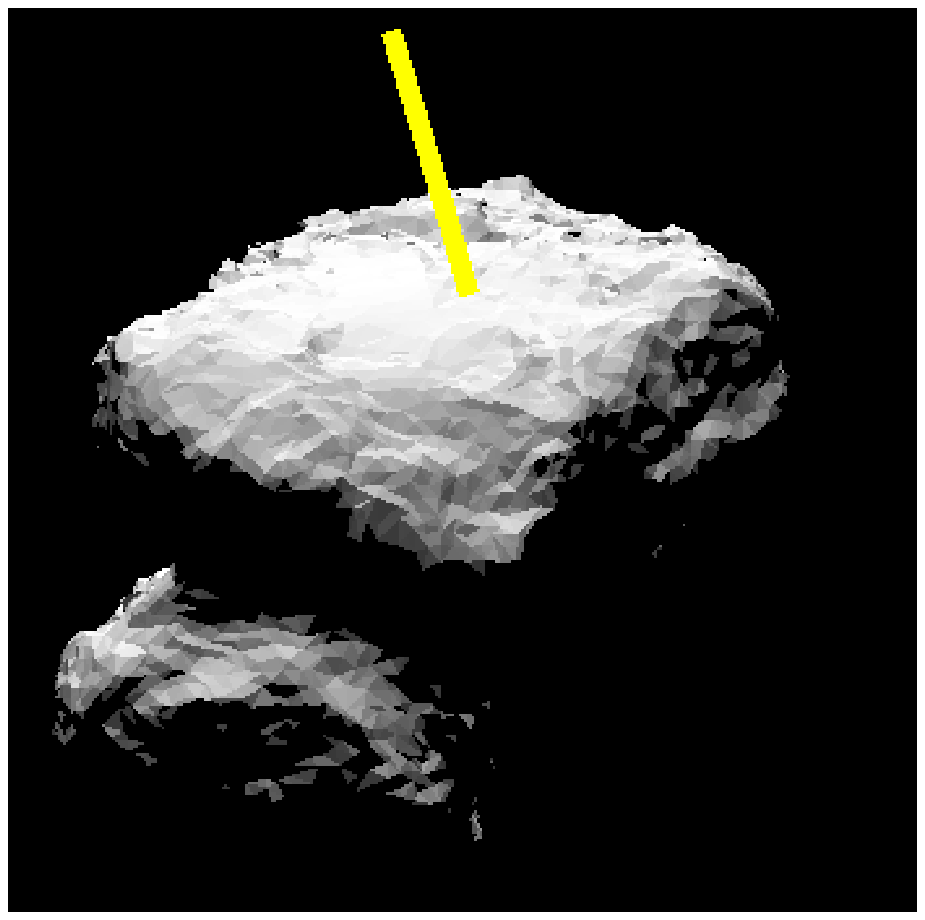}
  \vspace{0.3in}
\caption{Tri-dimensional simulation of the nucleus 
and the reconstructed jet at the time of observation (2015-05-30T17:13:38).}
\label{Imhotep_jet}
\end{figure}

			We determine the source of the jet using the technique described in 
\cite{Vincent_2016}. We identify the same jet structure in two images 
separated by one hour, i.e. 30 degrees of nuclear rotation 
(2015-05-30T17:13:38 \& 2015-05-30T18:13:39). It gives us a very good 
stereographic view of the jet, which allows for accurate triangulation 
of its source. We found that the feature on May 30, 2015 is most likely arising from a 
source located on the eastern side of the Imhotep region (see Figure \ref{Map_Thomas}). More exactly, it is located around latitude -$8^{\circ}$ and longitude 160$^{\circ}$ in the standard 
"Cheops" frame \citep{Preusker_2015}. The inversion defines the jet as a 
single line at the center of the feature. The uncertainty on the 
orientation of this line leads to a maximum error of about 10$^{\circ}$ in 
latitude and longitude for the source. The maximum error is estimated by looking at 
the scatter of the solution when repeating the inversion multiple times 
for slightly different definitions of the jet axis (geometric axis or 
brightest line for instance). The jet is pointing slightly away from the observer, and the exact angle is changing for each image.
The source is extremely close to the subsolar latitude (-8.8$^{\circ}$) 
which is consistent with previous studies showing that activity is 
generally following the subsolar latitude \citep{Vincent_2016}. This area is 
also interesting for being at the boundary between the edge of 
the smooth Imhotep plains and rougher terrains where evidence of surface water 
ice has been found \citep{Auger_2015, Oklay_2016, Pommerol_2015, Filacchione_2015, Barucci_2016}. Figure 
\ref{Imhotep_jet} shows a simulation of the nucleus 
and the reconstructed jet at the time of observation (2015-05-30T17:13:38).\\
	
\section{Model of sublimation of icy aggregates}
\label{Model_subli}
We aim to analyze the sublimation of outflowing, icy aggregates by considering an ensemble of aggregates of various radii and different compositions. 
In this work, three grain compositions are considered: pure water ice, a homogeneous mixture of dirty ice, and two-layer grains. Pure water ice is the most extreme case, while layered and mixed grains represent two different models of pebble formation in the protoplanetary disk. Layered grains should reflect the radial transport of refractory material accreted in the warm inner disk to the cold outer part, where frozen volatiles should form a mantle on the surface of the grains. By contrast, an intimate mixture of refractories and volatiles throughout the grains would suggest that the refractories accreted beyond the snowline of the protoplanetary disk, or that aggregates transported outward were so fluffy that gas could enter the pores inside them. Also, these two grain models are standard models of cometary dust, e.g. \cite{Greenberg_Hage_1990} for the layered grains. \\

The aggregates are formed from submicron-sized grains. The indices of refraction $m$ are complex and wavelength-dependent parameters ($m$  = $n$ - i$k$). In the pure water ice case, we use $n$ and $k$ given by Warren (1984). The dirty grains are $\rm H_2O$ ice mixed with ammonia and amorphous carbon, for which we used the artificial $n$ and $k$ given by \cite{Preibisch_1993}. In the two-layer case, we considered a silicate core \citep[olivine with Mg = Fe = 0.5,][]{Harker_2007, Gicquel_2012} covered by a water ice mantle. The indices $m$ that we use for the amorphous silicates are from \cite{Dorschner_1995}. \\

As described by \cite{Greenberg_Hage_1990} and \cite{ Gicquel_2012}, we used the Maxwell Garnett formula to calculate $m_{bi}$ of a two-layer grains. The indices of refraction depend on the fractional mass of the mantle component. We adopt the density $\rho_{Si}$ = 3.3 g $\rm cm^{-3}$ for silicate grains and $\rho_{W}$ = 1.0 g $\rm cm^{-3}$ for water ice and dirty grains \citep{Harker_2002, Gicquel_2012}. The fractional mass of the water ice mantle encasing the silicate core, $\alpha_{ice}$ is set to 0.2 for the two-layer grains. In the case of the two-layer grains, $\alpha_{ice}$ is constrained \citep{Rotundi_2015} with a dust/gas mass ratio of 4 $\pm$ 2 at $\rm R_h$ between 3.6 and 3.4 AU \footnote{\cite{Rotundi_2015} combined the data from the Grain Impact Analyser and Dust Accumulator (GIADA) with OSIRIS, Microwave Instrument for the Rosetta Orbiter (MIRO) and Rosetta Orbiter Spectrometer for Ion and Neutral Analysis (ROSINA) instruments.}. Also, based on on their model and combined with water production rates from MIRO, \cite{Moreno_2016} derived a 3.8 $<$ dust-to-gas mass ratio $<$ 6.5.\\

As discussed by \cite{Gicquel_2012}, we used another form of the Maxwell Garnett equation to calculate the refractive index $ m_p$ of porous aggregates. Aggregates are composed of submicron grains as defined by \cite{Greenberg_Hage_1990}. The radius of the submicron grains is 10 $\%$ of the radius of the aggregates. For the purpose of this article, we fix the porosity ($p$) equal to 0.5 which is in the range of values obtained with the COmetary Secondary Ion Mass Analyser COSIMA \citep{Schulz_2015}, and the Dust Impact Monitor DIM \citep{Kruger_2015}.\\

As described in \cite{Gicquel_2012}, the equilibrium temperature
of the aggregates ($T_d$) is derived by solving a thermal equilibrium (absorption from the Sun, radiation in the
infrared and cooling by sublimation). We use Mie theory \citep{Bohren_1983, Van_de_Hulst_1957} to compute the scattering properties of an assumed aggregate (size, composition, and porosity). \\

\begin{figure}
\centering
\includegraphics[scale=0.30]{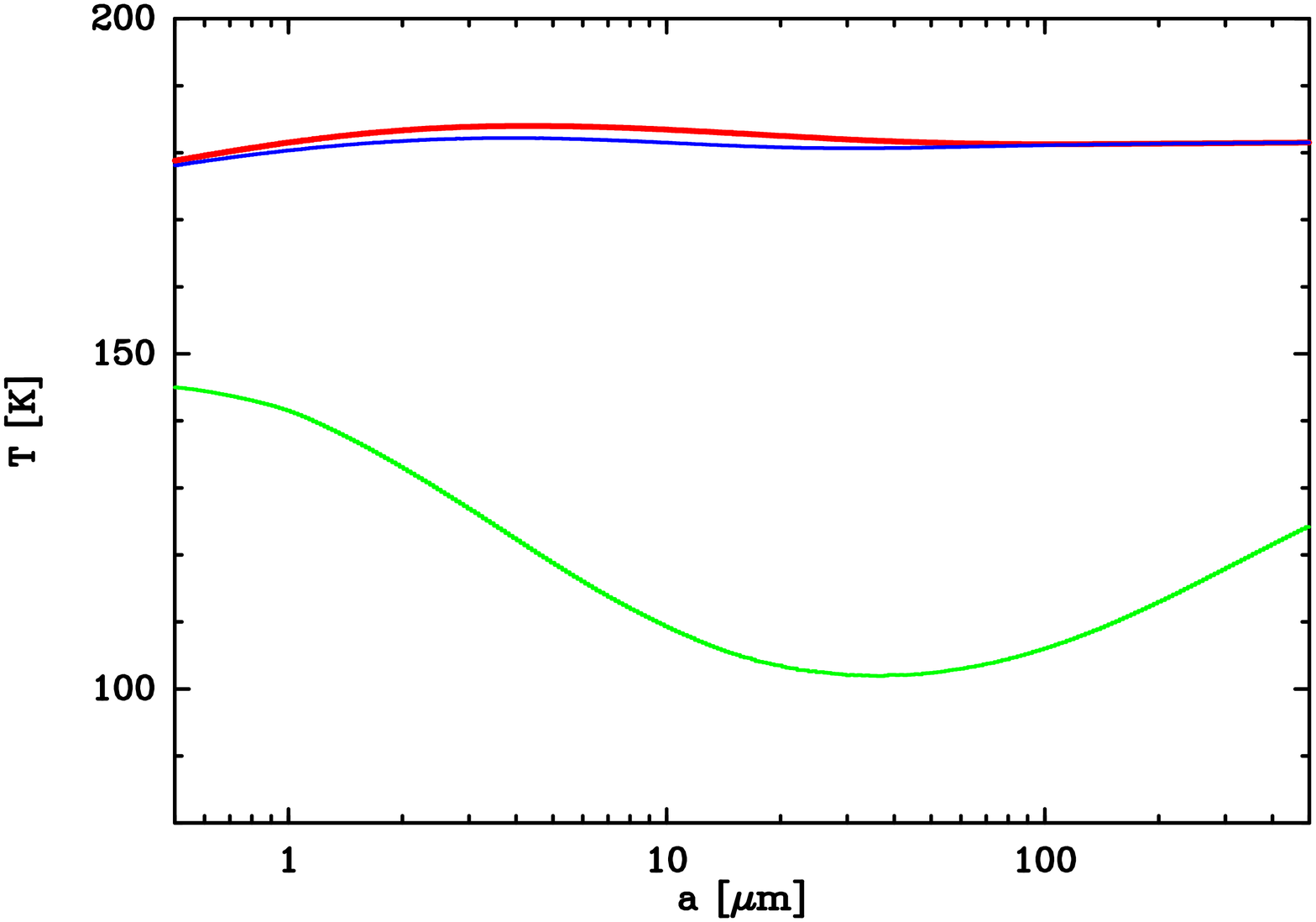}
  \vspace{0.3in}
\caption{The equilibrium temperature of aggregates for the three grain composition as a function of the aggregates radius: dirty ice (red), water ice (green), and two-layer grains (blue) at $\rm R_h$ = 1.53 AU. The porosity is 0.5 in all cases. }
\label{Temperature}
\end{figure}

Figure \ref{Temperature} shows the equilibrium temperature of the aggregates of two-layer grains, dirty ice and water ice at the heliocentric distance of our observations ($\rm R_h$ = 1.53 AU,)  as a function of the aggregate's radius ($a$). The aggregates radii vary from 0.5$\rm \mu$m to 500$\rm \mu$m, which is in the same range as the ones detected by the GIADA instrument \citep{Della_Corte_2015, Fulle_2015}. As we show in Section \ref{jet-param}, these turn out to be the allowed extremes for fitting the data. As asserted by \cite{Gunnarsson_2003, Beer_2006} and \cite{Gicquel_2012}, the temperature of aggregates is not particularly sensitive to the exact amount of impurities within the ice. However, the difference from pure water ice aggregates is important. \\

As shown in \cite{Gicquel_2012}, once ejected into the coma, the evolution of the aggregates depends on their physical properties (size, composition) and the geometry of the observations (heliocentric distance, because the temperature of
the aggregates are higher close to the Sun). We follow the evolution of aggregates until they stop sublimating. The sublimation results in a temporal variation of the aggregate's radius given by:
\begin{equation}
	\left|\frac{da}{dt}\right|=\frac{P_v(T_{d})}{\rho_d}\sqrt{\frac{ m_{H_{2}O}}{2 \pi k_B T_{d}}} 
	\label{dadt}
  \end{equation}
	where $k_B$ is
the Boltzmann constant, $ m_{\rm H_{2}O}$ is the mass of the water molecule and $\rho_d$ is the density of the dust. We used the vapor pressure, $P_v$, derived from \citet{Lichtenegger_1991} as expressed in \cite{Gicquel_2012}:

\begin{equation}
P_v\left(T_d\right)=P_r exp\left[\frac{m_{H_{2}O} L}{k_B}  \left(\frac{1}{T_{r}} - \frac{1}{T_{d}}\right)\right]
\label{FormulePression}
  \end{equation}

where $P_r$ = $\rm 10^5$ N $\rm m^{-2}$, $T_r$ = 373 K and L = 2.78 $\times$ $10^6$ J $\rm kg^{-1}$.\\

The mass loss (kg $\rm s^{-1}$) due to the sublimation, is expressed as:
\begin{equation}
\frac{dm_{ice}}{dt} = - 4 \pi a^2 S_A P_v\left(T_{d}\right)\sqrt{\frac{ m_{H_{2}O}}{2 \pi k_B
T_{d}}} \label{dadt}
  \end{equation}

where $S_A$ is the correcting factor, reflecting the ratio of the surface area of the submicrons grains into the total sublimating area of the porous aggregates. The equations used in this article are given by \cite{Gunnarsson_2003}. The factor $S_A$ is based only on the assumption that aggregates are composed of smaller component spheres. The reader is referred to \cite{Gunnarsson_2003} for a complete explanation of this parameter.\\

The aggregate lifetime is the time needed for the ice of aggregate to sublimate completely. Given existing suppositions on the velocity of icy aggregates velocity (Eq. \ref{vel}), we can determine at which distance from the nucleus the sublimation starts and ends. The factor $S_A$ influences directly the aggregate lifetime. If we suppose that the grains are stuck together or that the aggregates have a
lesser exposed surface area, we would derive a longer lifetime. \\

For comet 67P, the terminal velocity of the aggregates (m $\rm s^{-1}$) after they leave the surface can be approximated by the scaling law \citep{Vincent_2013, Fulle_1987, Finson_Probstein_1968} :
\begin{equation}
v =v_0\left(a\right) \left(\frac{4 \times 10^{-7}}{2a}\right)^{1/6}
\label{vel}
  \end{equation}
	where $v_0\left(a\right)$ is the initial velocity of the aggregates as a function of the initial radius of the aggregates. The velocity law is calibrated using the measurements by GIADA \citep{Rotundi_2015, Della_Corte_2015} and ground-based observations \citep{Vincent_2013} : $v_0\left(5\mu m \right)$= 80 m $\rm s^{-1}$, $v_0\left(50\mu m\right)$= 30 m $\rm s^{-1}$ and $v_0\left(500\mu m\right)$= 10 m $\rm s^{-1}$. In this study we aim to explain the curve in the radial profile of the jet from 4 to 10 km. \cite{Lin_2015} show a change in slope due to the acceleration of the dust in the innermost coma (up to 2 km). However, we don't see this signature in our data (Figure \ref{Vis_Jet_Coma}). It might be because the source region of the jet is not in the FOV. Therefore, we assume that the acceleration of the aggregates is negligible for the purpose of this paper. The reader is referred to \cite{Agarwal_2016} for a more detailed investigation of the acceleration of aggregates within 2 km from the surface.

\section{Retrieving the jet parameters}
\label{jet-param}
\subsection{Results of model simulations}
\label{Model_parameter}

We present the evolution of the aggregates as a function of the composition at the heliocentric distance of our observations. We will not consider pure water ice aggregates. As an example, at $\rm R_h$ = 1.53 AU, pure water ice aggregates with a radius $a$ = 5$\rm \mu$m stop sublimating at $D$ > $\rm 10^5$km to far outside our field of view to be meaningful. We consider aggregates of two-layer grains and dirty grains with a radius at $a$ = 5$\rm \mu$m, 50$\rm \mu$m and 500$\rm \mu$m. Table \ref{Compo} summarizes at which distance from the nucleus the aggregates of dirty and two-layer grains stop sublimating. We can see that the sublimation in the case of aggregates of two-layer grains is more efficient. Indeed, the aggregates of dirty grains stop sublimating further from the nucleus than the aggregates of two-layer grains. Even if the temperature of aggregates that have 'dirt' contamination is similar (Figure \ref{Temperature}), the amount of ice in the case of the aggregates of dirty grains is higher than for aggregates of two-layer grains. As an example, for aggregates with a radius at $a$ = 50$\rm \mu$m, the aggregates stop sublimating at $D$ around 4 km and 10 km for the two-layer grains and the dirty grains, respectively. For $a$ = 500$\rm \mu$m, the aggregates stop sublimating at $D$ around 7 km and 25 km for the two-layer grains and the dirty grains, respectively. \\

\begin{table}
\caption{Distance from the nucleus where the aggregates of dirty and two-layer grains stop sublimating for various radii at $\rm R_h$ = 1.53 AU.}
\label{Compo}
\begin{tabular}{lll}
  \hline
  a& Composition & D  \\
 ($\mu$m) & & (km) \\
  \hline
5	& 2-Layer & 1.1 \\
& Dirty & 2.8 \\
50& 2-Layer &  3.6\\
& Dirty & 10.5 \\
500& 2-Layer & 7.1 \\
& Dirty & 23.7 \\

\hline
\end{tabular}
\end{table}

As noted earlier, the identical radial profiles with the Visible and UV filters highlights that the composition does not change, which means that the grains may be a mixture rather than layered. Also, the pure water ice aggregates sublimate too far away from the nucleus. For these reasons, only the radial profiles of the jet obtained with the dirty aggregates will be compared with the OSIRIS data. Also, Table \ref{Compo} allows us to reduce the size range of the dirty aggregates. As described in Section \ref{Model_subli}, we first considered the sublimation of the aggregates of dirty grains with a radius from 0.5$\mu$m to 500$\mu$m. However, with a radius $a$ $>$ 5$\mu$m and $<$ 50$\mu$m, the aggregates stop sublimating at a distance from the nucleus $D$ $>$ 1 km and $<$ 10 km. As explained in Section \ref{Rad_Pro}, we did not consider the radial profile in the inner coma to avoid possible straylight contamination from the nucleus, and the jet brightness was below the background noise after 10 km. It does not exclude that sublimation exists at $D$ $<$ 1 km and $>$ 10 km, but for the data we used, only sizes between 5$\mu$m and $<$ 50$\mu$m are relevant.


\subsection{Comparison with the OSIRIS images}
\label{Model_Comparaison}

\begin{figure*}
\centering
\includegraphics[scale=0.40]{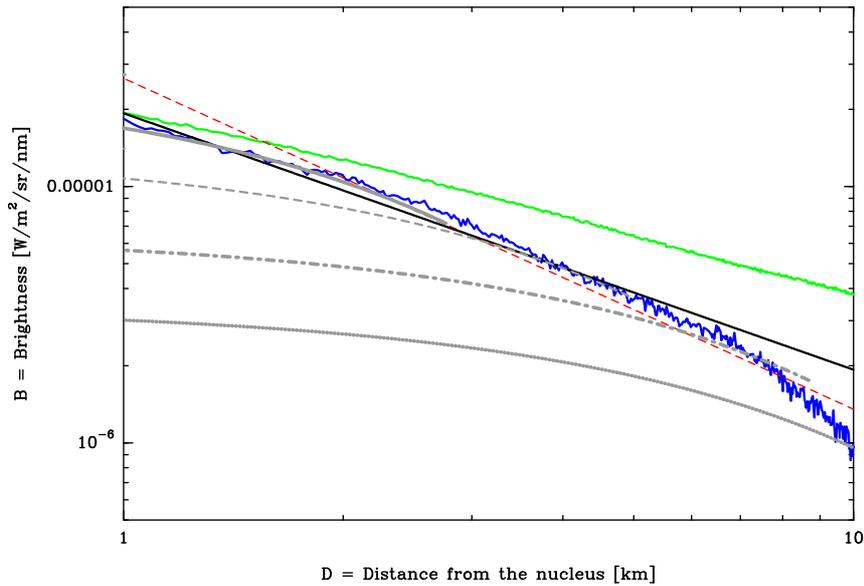}
  \vspace{0.3in}
\caption{2015-05-30T16:43:38 - In blue is the radial profile for the jet and in green are the radial profile for the coma background. In red is the $D^{\beta}$-fit for the jet. In grey are the result of sublimation of the aggregates of dirty grains with a radius at a = 5$\mu$m (Solid Line), a = 10$\mu$m (Dashed Line), a = 20$\mu$m (Dash-Dotted Line) and a = 50$\mu$m (Dotted Line). In black is the radial profile over a cone. }
\label{Method_Plot}
\end{figure*}

As shown in Figure \ref{Method_Plot}, we focus on the radial brightness of the jet and the background coma on 2015-05-30T16:43:38 to explain the curve from D = 4 km to 10 km. We first thought that the curve beyond 4 km was due to the dispersion of the jet over a cone. However, the dispersion, as a function of the distance from the nucleus, reproduces the brightness until 4 km but not after that (Figure \ref{Method_Plot}, in black). Actually, the jet on May 30, 2015 seems to be collimated, and it does not follow a divergent pattern. We consider the sublimation of the aggregates of dirty grains with a radius from 5$\mu$m to 50$\mu$m, which is motivated by the results in Table \ref{Compo}. We can see that with $a$ = 5$\mu$m, 10$\mu$m, 20$\mu$m and 50$\mu$m, we are able to fully reproduce the curve of the radial brightness of the jet from D = 4 km to 10 km. The result of the comparison between the model and the OSIRIS data is a strong constraint of the icy particle sizes, and gives us an indication of the number of aggregates ($N_{aggregate}$) we need to reproduce the observed brightness flux, $B$, in the OSIRIS images. However, other intimate mixtures with different radii might have been able to reproduce the curve of the radial brightness of the jet.  \\

We suppose that the theoretical brightness for a single aggregate $I$ (W $\rm m^{-2}$ $\rm nm^{-1}$ ) is expressed as:
\begin{equation}
	I=\frac{A\phi\left(\alpha\right)}{\pi} \frac{F_{Sun,\lambda_{VIS}}}{R_h^2}\frac{1}{\Delta_{S/C}^2}  \pi a^2
\end{equation}
where $A$ is the geometric albedo, $\phi$ is the phase function, $\alpha$ is the phase angle, and $F_{Sun,\lambda_{VIS}}$ is the flux of the Sun at the central wavelength of the visible filter. With a phase angle $\alpha$ = 70 deg (Table \ref{Obs_VISIBLE}), we derive a phase function $\phi(70)$ = 0.04 \citep{Fornasier_2015}, and we suppose an albedo $A$ = 6.5 $\times$ $10^{-2}$.\\

The number of aggregates per pixel we need to reproduce the observed brightness flux (Figure \ref{Method_Plot}) is: $N_{px}$ = $B$ $A_{px}$ /$I$, where $A_{px}$ is the solid angle of a single pixel. The aggregates contain a starting mass of $\rm H_2O$ ice (kg), given by: $M_{ice}$ = (4/3) $\pi$ $a^3$ $\rho_{W}$ (1-$p$). The values of $N_{aggregate}$ and $M_{ice}$ as a function of the size of the aggregates are given in Table \ref{Size_Nb}. To reproduce the data we need $N_{aggregate}$ between 8.5 $\times$ $10^{13}$ and 8.5 $\times$ $10^{10}$ for a = 5$\mu$m and 50$\mu$m, respectively. The total number of aggregates correspond to $M_{ice}$ $\approx$ 22kg. The total number of aggregates with a radius at $a$ = 50$\mu$m is in good agreement with the total number of particles with a radius at $a$ = 50$\mu$m in the coma up to 15 km from 2 AU to perihelion \citep{Fulle_2016}. The dust loss rate corresponding to the observed $N_{aggregate}$ within 15 km is about 0.1 kg $\rm sec^{-1}$, compared
to a total dust loss rate of 70 kg $\rm sec^{-1}$ at 2 AU, and of 1500 kg $\rm sec^{-1}$ at perihelion \citep{Fulle_2016}.

\begin{table}
\caption{$N_{aggregate}$ and $M_{ice}$ for each size of the aggregates to reproduce the radial profile of the jet.}
\label{Size_Nb}
\begin{tabular}{lll}
  \hline
 a  & $N_{aggregate}$ & $M_{ice}$  \\
  ($\mu$m)& & (kg) \\
  \hline
 5& 8.5 $\times$ $10^{13}$ & 22.4 \\
10& 1.1 $\times$ $10^{13}$ & 22.4 \\
20& 1.3 $\times$ $10^{12}$ & 22.4 \\
50 &8.5 $\times$ $10^{10}$  & 22.4 \\
\hline
\end{tabular}
\end{table}

\section{Discussion and conclusion}

We used WAC images from the OSIRIS cameras with the Visible filter shortly after the southern vernal equinox on May 30 and 31, 2015 at $\rm R_h$ = 1.53 AU (Table \ref{Obs_VISIBLE}), as explained in Section \ref{Obs_Section}. We studied the temporal variation of the brightness of the jet and the background coma and found that the jet rotates with the nucleus (Figure \ref{Image_Jet_Coma}). The key result of this work is the determination of the brightness slopes of the jet (1.31 $\pm$ 0.17) which is much steeper than that of the background coma (0.74 $\pm$ 0.06), as determined in Section \ref{Rad_Pro} (Figure \ref{Vis_Jet_Coma}). This is a first indication of sublimation as also suggested by \cite{Hofstadter_2016} with MIRO. From OSIRIS observations at $\rm R_h$ = 3.5 AU (from August to September 2014), \cite{Lin_2015} found similar results for both the slope for the background and for the jet. Even if the heliocentric distance is different, the similar results between these studies are remarkable. Nonetheless, more data analysis on the OSIRIS images has to be done to fully understand the composition and morphology of the jets. Furthermore, the slope we derived for the background coma was consistent with results from various instruments on comet 67P \citep{Lin_2015, Lara_2011, Hadamcik_2010} and 103P/Hartley 2 \citep{Protopapa_2014}. \\

We determined, in Section \ref{Jet_Location}, that the jet arose from the eastern side of the Imhotep region. The source of the jet we are studying was at the boundary between the smooth edge of the Imhotep plains and rougher terrains where evidence of water ice on the surface havd been found (Section \ref{Jet_Location}, Figure \ref{Imhotep_jet}). Bright features in the Imhotep region, which might be indicative of ice and water ice in particular \citep{Oklay_2016,Pommerol_2015}, have been detected on the Imhotep region with OSIRIS \citep{Groussin_2015, Auger_2015}, and VIRTIS \citep{Filacchione_2015, Barucci_2016}. The bluer regions looked exceptionally active with jets \citep{Vincent_2016, Oklay_2016} or outbursts rising from the bright patches. All these studies confirmed the presence of water ice on the surface and subsurface of comet 67P (and probably indicated a mixture of ice and dust). \\

In Section \ref{Model_subli}, we described an updated model of sublimation of icy aggregates \citep{Gicquel_2012}. As explained in the present paper, even if the temperature of the aggregates with a 'dirt' contamination was similar (Figure \ref{Temperature}), the amount of ice played an important role in the lifetime of the aggregates of icy grains and at which distance from the nucleus the sublimation occured. Table \ref{Compo}, in Section \ref{Model_parameter}, allowed us to constrain both the composition (dirty grains) and the size (radius 5$\mu$m - 50$\mu$m) of the icy aggregates. In the study of the jet, observed near the southern vernal equinox on May 2015, we supposed a length scale of the jet around 10 km \citep{Lin_2015, Tozzi_2004}. The jet brightness is below the background noise after 10 km. In Section \ref{Model_Comparaison}, we compared the model with the OSIRIS images. The major result of this article is that by sublimation of the aggregates of dirty grains, we can explain the decline beyond 4km from the nucleus in the radial brightness profile of a jet (Section \ref{Model_Comparaison}, Figure \ref{Method_Plot}). However, other intimate mixtures with different radii might be able to reproduce the curve of the radial brightness of the jet. To reproduce the data we needed $N_{aggregate}$ between 8.5 $\times$ $10^{13}$ and 8.5 $\times$ $10^{10}$ for a = 5$\mu$m and 50$\mu$m respectively, or an initial mass of $\rm H_2O$ ice around 22kg (Table \ref{Size_Nb}). Our total number of aggregates of radius 50 micron is in good agreement with \cite{Fulle_2016}. \\

\cite{Belton_2010} proposed a nomenclature of mechanisms for producing jets. His Type I jet appeared to be related to the sublimation of water with the release of dust due to the sublimation of icy grains below the surface. The COSIMA instrument, aimed to collect comet particles \citep{Kissel_2007} They reported both the detections of particles with a radius $>$ 50$\mu$m and a porosity much higher that the porosity of the nucleus at $\rm R_h$ = 3.57 AU  \citep{Schulz_2015}. Although no ice was detected, it could indicate that the ice component already sublimated before the COSIMA images were taken. If the ice was below the surface, we needed efficient gas drag to remove the dust which covered the surface for many years. However, closer to the Sun the volatiles underneath the dust layers could lift the overlying dust. Our work showed that coordination of Rosetta data, ground-based observations and numerical models will help to have a better understanding of the jet's activity. However, the mechanisms creating the jets is complex and will need further analysis. Furthermore, we need to compare the radial profile of jets from others regions of the nucleus with our model to have a better understanding of the structure and composition of the nucleus. \\

\section*{Acknowledgements}

OSIRIS was built by a consortium of the Max-Planck-
Institut f{\"u}r Sonnensystemforschung, G{\"o}ttingen, Germany, CISAS University of Padova, Italy, the Laboratoire d'Astrophysique de Marseille, France, the
Institutode Astrof\'isica de Andalucia, CSIC, Granada, Spain,the Research and
Scientific Support Department of the European Space Agency, Noordwijk, The
Netherlands, the Instituto Nacionalde T\'ecnica Aeroespacial, Madrid, Spain, the
Universidad Polit\'echnica de Madrid, Spain, the Department of Physics and
Astronomy of Uppsala University, Sweden, and the Institut f{\"u}r Datentechnik und
Kommunikationsnetze der Technischen Universit{\"a}t Braunschweig, Germany.
The support of the national funding agencies of Germany (DLR), France(CNES),
Italy(ASI), Spain(MEC), Sweden(SNSB), and the ESA Technical Directorate
is gratefully acknowledged. We thank the Rosetta Science Ground Segment at
ESAC, the Rosetta Mission Operations Centre at ESOC, and the Rosetta Project
at ESTEC for their outstanding work enabling the science return of the Rosetta
Mission.


\bsp	
\label{lastpage}

\begin{thebibliography}{}

\bibitem[Agarwal et al.(2016)]{Agarwal_2016}{Agarwal}, J., {A'Hearn}, M.~F., {Vincent}, J.-B., et al. 2015, MNRAS, this issue

\bibitem[A'Hearn et al.(2011)]{AHearn_2011}{A'Hearn}, M.~F., {Belton}, M.~J.~S., {Delamere}, W.~A., et al. 2011, Science, 332, 1396

\bibitem[Auger et al.(2015)]{Auger_2015}{Auger}, A.-T., {Groussin}, O., {Jorda}, L., et al. 2015, A\&A, 583, A35

\bibitem[Barucci et al.(2016)]{Barucci_2016}Barucci, M.-A, Filacchione, G., Fornasier, S. et al. 2016, A\&A, submitted

\bibitem[Beer et al.(2006)]{Beer_2006}{Beer}, E.~H., {Podolak}, M., {Prialnik}, D. 2006, Icarus, 180, 473

\bibitem[Belton(2010)]{Belton_2010}Belton, M.~J.~S. 2010, Icarus, 210, 881

\bibitem[Bohren \& Huffman(1983)]{Bohren_1983}Bohren, C. F. \& Huffman, D. R. 1983, New York: Wiley

\bibitem[Capaccioni et al.(2015)]{Capaccioni_2015}{Capaccioni}, F, {Coradini}, A., {Filacchione}, G., et al. 2015, Science, 347, aaa0628

\bibitem[Choukroun et al.(2015)]{Choukroun_2015}{Choukroun}, M., {Keihm}, S., {Schloerb}, F.~P., et al. 2015, A\&A, 583, A28


\bibitem[De Sanctis et al.(2015)]{DeSanctis_2015}{De Sanctis}, M.~C. and {Capaccioni}, F. and {Ciarniello}, M., et al. 2015, Science, 525, 500

\bibitem[Della Corte et al.(2015)]{Della_Corte_2015}{Della Corte}, V., {Rotundi}, A., {Fulle}, M., et al. 2015, A\&A, 583, A13

\bibitem[Dorschner et al.(1995)]{Dorschner_1995}Dorschner, J., Begemann, B., Henning, T., et al. 1995, A\&A, 300, 503

\bibitem[Farnham et al.(2013)]{Farnham_2013}{Farnham}, T.~L., {Bodewits}, D., {Li}, J.-Y., et al. 2013, Icarus, 222, 540

\bibitem[Feaga et al.(2007)]{Feaga_2007}{Feaga}, L.~M., {A'Hearn}, M.~F., {Sunshine}, J.~M., et al. 2007, Icarus, 190, 345

\bibitem[Filacchione et al.(2016)]{Filacchione_2015}{Filacchione}, G., {de Sanctis}, M.~C., {Capaccioni}, F., et al. 2016, Nature, 529, 368

\bibitem[Finson \& Probstein(1990)]{Finson_Probstein_1968}Finson, M. L., \& Probstein, R. F. 1968, ApJ, 154, 353

\bibitem[Fornasier et al.(2015)]{Fornasier_2015}Fornasier, S., Hasselmann, P. H., Barucci, M. A., et al. 2015, A\&A, 583, A30

\bibitem[Fulle(2007)]{Fulle_1987}Fulle, M. 1987, A\&A, 171, 327

\bibitem[Fulle et al.(2015)]{Fulle_2015}Fulle, M., {Della Corte}, V., {Rotundi}, A., et al. 2015, ApJL, 802, L12

\bibitem[Fulle et al.(2016)]{Fulle_2016}Fulle, M., {Marzari}, F., {Della Corte}, V., et al. 2016, ApJL, A\&A, 821, 19

\bibitem[Gicquel et al.(2012)]{Gicquel_2012}{Gicquel}, A., {Bockel{\'e}e-Morvan}, D., {Zakharov}, V.V, et al. 2012, A\&A, 542, A119

\bibitem[Gunnarsson(2003)]{Gunnarsson_2003}{Gunnarsson}, M. 2003, A\&A, 398, 353

\bibitem[Greenberg \& Hage(1990)]{Greenberg_Hage_1990}{Greenberg}, J.~M. \& {Hage}, J.~I. 1990, ApJ, 361, 260
 
\bibitem[Groussin et al.(2015)]{Groussin_2015}{Groussin}, O., {Sierks}, H., {Barbieri}, C., et al. 2015, A\&A, 583, A36

\bibitem[Gr{\"u}n et al.(2016)]{Grun_2016}{Gr{\"u}n}, E., {Agarwal}, J., {Altobelli}, N., et al. 2015, MNRAS, this issue

\bibitem[Hadamcik et al.(2010)]{Hadamcik_2010}{Hadamcik}, E., {Sen}, A.~K., {Levasseur-Regourd}, A.~C., et al. 2010, A\&A, 517, A86

\bibitem[Hanner \& Zolensky(2010)]{Hanner_Zolensky_2010}{Hanner}, M.~S. \& {Zolensky}, M.~E. 2010, LNP, 815, 203

\bibitem[Harker et al. (2007)]{Harker_2007}Harker, D. E.,Woodward, C. E., Wooden, D. H., et al. 2007, Icarus, 190, 432

\bibitem[Hofstadter et al.(2016)]{Hofstadter_2016}{Hofstadter}, M., {Schloerb}, F.~P., {Gulkis}, S., et al. 2016, 50th ESLAB Symposium, P31

\bibitem[Harker et al.(2002)]{Harker_2002}{Harker}, D.~E., {Wooden}, D.~H., {Woodward}, C.~E., et al. 2002, ApJ, 580, 579

\bibitem[Haser(1957)]{Haser_1957}{Haser}, L. 1957, Bulletin de la Societe Royale des Sciences de Liege, 43, 740

\bibitem[Keller et al.(2007)]{Keller_2007}Keller, H. U., Barbieri, C., Lamy, P., et al. 2007, SSRv, 128, 433

\bibitem[Keller et al.(2015)]{Keller_2015}Keller, H. U., Mottola, S., {Davidsson}, B., et al. 2015, A\&A, 583, A34

\bibitem[Kissel et al.(2007)]{Kissel_2007}{Kissel}, J., {Altwegg}, K., {Clark}, B.~C., et al 2007, SSRv, 128, 823

\bibitem[Knollenberg et al.(2016)]{Knollenberg_2016}Knollenberg, J., {Lin}, Z.-Y., Hviid, S.~F., et al. 2016, A\&A, submitted

\bibitem[Lara et al.(2011)]{Lara_2011}{Lara}, L.~M., {Lin}, Z.-Y., {Rodrigo}, R., et al. 2011, A\&A, 526, A36

\bibitem[Lara et al.(2015)]{Lara_2015}{Lara}, L.~M., {Lowry}, S., {Vincent}, J.-B., et al. 2015, A\&A, 583, A9

\bibitem[Li et al.(2013)]{Li_2013}Li, J.-Y., Besse, S., A'Hearn, M. F., et al. 2013, Icarus, 222, 559

\bibitem[Lichtenegger \& Komle(1991)]{Lichtenegger_1991}{Lichtenegger}, H.~I.~M. \& {Komle}, N.~I. 1991, Icarus, 90, 319

\bibitem[Lin et al.(2015)]{Lin_2015}{Lin}, Z.-Y., {Ip}, W.-H., {Lai}, I.-L., et al. 2015, A\&A, 583, A11

\bibitem[Lin et al.(2016)]{Lin_2016}{Lin}, Z.-Y., {Lai}, I.-L., {Su}, C.-C., et al. 2016, A\&A, 588, L3 

\bibitem[Lisse et al.(2006)]{Lisse_2006}Lisse, C.~M., {VanCleve}, J., {Adams}, A.~C., et al. 2006, Science, 313, 635 

\bibitem[Kr{\"u}ger et al.(2015)]{Kruger_2015}{Kr{\"u}ger}, H., {Seidensticker}, K.~J., {Fischer}, H.-H., et al. 2015, A\&A, 583, A15

\bibitem[Moreno et al.(2016)]{Moreno_2016}{Moreno}, F., {Snodgrass}, C., {Hainaut}, O., et al. 2016, A\&A, 587, A155

\bibitem[Mottola et al.(2014)]{Mottola_2014}{Mottola}, S. and {Lowry}, S. and {Snodgrass}, C., et al. 2014, A\&A, 569, L2

\bibitem[Mueller et al.(2013)]{Mueller_2013}{Mueller}, B.~E.~A. and {Samarasinha}, N.~H. and {Farnham}, T.~L., et al. 2013, Icarus, 222, 799

\bibitem[Oklay et al.(2016)]{Oklay_2016}{Oklay}, N., {Vincent}, J.-B., {Fornasier}, S., et al. 2016, A\&A, 586, A80

\bibitem[Pommerol et al(2015a)]{Pommerol_2015}{Pommerol}, A., {Thomas}, N., {El-Maarry}, M.~R., et al. 2015a, A\&A, 583, A25

\bibitem[Pommerol et al(2015b)]{Pommerol_2015b}{Pommerol}, A., {Jost}, B., {El-Maarry}, M.~R., et al. 2015b, P\&SS, 109, 106

\bibitem[Preibisch et al.(1993)]{Preibisch_1993}{Preibisch}, T., {Ossenkopf}, V., {Yorke}, H.~W. \& {Henning}, T. 1993, A\&A, 279, 577

\bibitem[Preusker et al.(2015)]{Preusker_2015}{Preusker}, F., {Scholten}, F., {Matz}, K.-D., et al. 2015, A\&A, 583, A33

\bibitem[Protopapa et al.(2014)]{Protopapa_2014}{Protopapa}, S., {Sunshine}, J.~M., {Feaga}, L.~M. et al. 2014, Icarus, 238, 191

\bibitem[Protopapa et al.(2015)]{Protopapa_2015}{Protopapa}, S., {Grundy}, W.~M., {Tegler}, S.~C. et al. 2015, Icarus, 253, 179

\bibitem[Rotundi et al.(2015)]{Rotundi_2015}{Rotundi}, A., {Sierks}, H., {Della Corte}, V. et al. 2015, Science, 347, a3905R

\bibitem[Schloerb et al.(2015)]{Schloerb_2015}{Schloerb}, F.~P., {Keihm}, S., {von Allmen}, P. et al, 2015, A\&A, 583, A29

\bibitem[Schulz et al.(2015)]{Schulz_2015}{Schulz}, R., {Hilchenbach}, M., {Langevin}, Y. et al, 2015, Nature, 518, 216

\bibitem[Shi et al.(2015)]{Shi_2015}{Shi}, X., {Hu}, X., {Sierks}, H., et al. 2016, A\&A, 586, A7

\bibitem[Sierks et al.(2015)]{Sierks_2015}{Sierks}, H., {Barbieri}, C., {Lamy}, P.~L. et al. 2015, Science, 347, a1044S

\bibitem[Sunshine et al.(2016)]{Sunshine_2006}Sunshine, J. M., A'Hearn, M. F., Groussin, O., et al. 2006, Science, 311, 1453

\bibitem[Thomas et al.(2015)]{Thomas_2015}Thomas, N., {Sierks}, H., {Barbieri}, C., et al. 2015, Science, 347, 6620 

\bibitem[Tubiana et al.(2015)]{Tubiana_2015}{Tubiana}, C., {Snodgrass}, C., {Bertini}, I., et al 2015,A\&A, 573, A62 

\bibitem[Tozzi et al.(2004)]{Tozzi_2004}{Tozzi}, G.~P., {Lara}, L.~M., {Kolokolova}, L., et al. 2004, A\&A, 424, 325

\bibitem[Tozzi et al.(2011)]{Tozzi_2011}Tozzi, G.~P., {Patriarchi}, P., {Boehnhardt}, H., et al. 2011, A\&A, 531, A54

\bibitem[Van de Hulst(1957)]{Van_de_Hulst_1957}Van de Hulst, H. C. 1957, Light Scattering by Small Particles, New York: John Wiley Sons

\bibitem[Vincent et al.(2010)]{Vincent_2010}{Vincent}, J.-B., {B{\"o}hnhardt}, H., {Lara}, L.~M. 2010, A\&A, 512, A60

\bibitem[Vincent et al.(2013)]{Vincent_2013}{Vincent}, J.-B., {Lara}, L.~M., {Tozzi}, G.~P., et al. 2013, A\&A, 549, A121

\bibitem[Vincent et al.(2016)]{Vincent_2016}{Vincent}, J.-B., {Oklay}, N., {Pajola}, M., et al. 2016, A\&A, 587, A14

\end{thebibliography}
\end{document}